\documentclass[a4paper,11pt]{article}

\usepackage{amsmath,amssymb,bbm}
\usepackage{ascmac}
\newdimen\tbaselineshift % for xy.sty with pdflatex
\usepackage{cancel}
\usepackage{xcolor} % for pdflatex
\usepackage{graphicx} % for pdflatex
\usepackage{fancybox}
\usepackage{shadow}
\usepackage{multirow}
\usepackage{marvosym,wasysym}
\usepackage{arydshln}
\usepackage{upgreek}
\usepackage[bookmarks=true,bookmarksnumbered=true,bookmarkstype=toc]{hyperref} 
\hypersetup{
pdfauthor={Tetsuji Kimura},
colorlinks={true},
linkcolor={blue},
urlcolor={blue},
filecolor={blue},
citecolor={blue}
}
\usepackage{ulem}

\makeatletter

 \@addtoreset{equation}{section}
\makeatother

\newcounter{Enumerate}

\DeclareFontFamily{U}{rsf}{}
\DeclareFontShape{U}{rsf}{m}{n}{
  <5> <6> rsfs5 <7> <8> <9> rsfs7 <10-> rsfs10}{}
\DeclareMathAlphabet\Scr{U}{rsf}{m}{n}
\usepackage[mathscr]{eucal}

\newcommand{\del}{\partial}
\newcommand{\half}{\frac{1}{2}}

\newcommand{\ls}{\ \ \ \ \ }
\newcommand{\wt}{\widetilde}
\newcommand{\wh}{\widehat}
\newcommand{\ve}{\varepsilon}
\newcommand{\ol}{\overline}

\newcommand{\bsubeq}{\begin{subequations}}
\newcommand{\esubeq}{\end{subequations}}

\newcommand{\noi}{\noindent}

\newcommand{\nn}{\nonumber}

\newcommand{\I}{{\rm i}}

\newcommand{\N}{\mathcal{N}}

\renewcommand{\d}{{\rm d}}
\newcommand{\e}{{\rm e}}

\newcommand{\slb}{\scalebox}

\def\+{{+\!\!\!+}}

\begin{document}
\allowdisplaybreaks{

\thispagestyle{empty}

%%%%%%%%% title %%%%%%%%%%%%%

\begin{flushright}
UTHEP-676, \
TIT/HEP-644 %\\
%[version 3.2] \\
%TK-NOTE/15-05 \\
%since: May 12, 2015 \\
%last update: %\today \\
%\timestamp
\end{flushright}

\vspace{35mm}

%%%%%%%%% title %%%%%%%%%
\noi
\slb{1.85}{Gauge-fixing Condition on Prepotential}

\vspace{5mm}

\noi
\slb{1.85}{of Chiral Multiplet for Nongeometric Backgrounds}
%%%%%%%%% title %%%%%%%%%

\vspace{15mm}

\noi
{\renewcommand{\arraystretch}{1.6}
\begin{tabular}{cl}
\multicolumn{2}{l}{\slb{1.2}{Tetsuji {\sc Kimura}}}
\\
& {\renewcommand{\arraystretch}{1.0}
\begin{tabular}{l}
{\sl Research and Education Center for Natural Sciences, Keio University}
\\
{\sl Hiyoshi 4-1-1, Yokohama, Kanagawa 223-8521, JAPAN} \footnotemark
%\\
%\slb{0.9}{\tt tetsuji.kimura \_at\_ keio.jp}
\end{tabular}
}
\\
& \ls and
\\
& {\renewcommand{\arraystretch}{1.0}
\begin{tabular}{l}
{\sl
Graduate School of Pure and Applied Sciences, 
University of Tsukuba} \vphantom{$\Big|$}
\\
{\sl Tsukuba, Ibaraki 305-8571, JAPAN}
%\\
%\slb{0.9}{\tt tetsuji \_at\_ het.ph.tsukuba.ac.jp}
\end{tabular}
}
\\
& \ls and
\\
& {\renewcommand{\arraystretch}{.8}
\begin{tabular}{l}
{\sl
Department of Physics,
Tokyo Institute of Technology} \vphantom{$\Big|$}
\\
{\sl Tokyo 152-8551, JAPAN}
%\\
%\slb{0.9}{\tt tetsuji \_at\_ th.phys.titech.ac.jp}
\end{tabular}
}
\\
& \ \ \ \slb{0.9}{\tt tetsuji.kimura \_at\_ keio.jp}
\end{tabular}
}
\footnotetext{The current affiliation since October 2015}

\vspace{10mm}
%\vfill

%%%%%%%%% abstract %%%%%%%%

\slb{1.1}{\sc Abstract}
\begin{center}
\slb{.95}{
\begin{minipage}{.95\textwidth}
\parindent=6mm%

We study a supergauge transformation of a complex superfield
which generates a chiral superfield in two-dimensional $\N=(2,2)$ theory.
This complex superfield is referred to as the prepotential of the chiral superfield.
Since there exist redundant component fields in the prepotential, 
we remove some of them by a gauge-fixing condition.
This situation is parallel to that of a vector superfield.
In order to obtain a suitable configuration of the GLSM for the exotic five-brane which gives rise to a nongeometric background, we impose a relatively relaxed gauge-fixing condition.
It turns out that the gauge-fixed prepotential 
is different from a semichiral superfield whose scalar field represents a coordinate of generalized K\"{a}hler geometry.
\end{minipage}
}
\end{center}

%%%%%%%%%%%%%%%%%%%%%%%%%%%%%%%%%%%%%%%%%%%%%%%%%%%%%%%%%%%%%%%%%%%%%%%
\newpage
\section{Introduction}
\label{sect:introduction}

Two-dimensional supersymmetric gauge theory is a powerful model when one studies string theory on curved geometry.
In particular, $\N=(2,2)$ gauge theory with charged matter plays a central role in analyzing string theory on Calabi-Yau manifolds and the corresponding Landau-Ginzburg theories \cite{Witten:1993yc}.
This gauge theory is called the gauged linear sigma model (GLSM).
Applying duality transformations of chiral superfields discussed in \cite{Rocek:1991ps} to the GLSM,
one can also study a supersymmetric sigma model on a Calabi-Yau manifold and its mirror dual geometry \cite{Hori:2000kt} via the Strominger-Yau-Zaslow (SYZ) conjecture \cite{Strominger:1996it}.

Two-dimensional gauge theory with eight supercharges, namely $\N=(4,4)$ GLSM, is also a significant tool for studying string propagation in the background geometry of NS5-branes and its T-dualized system \cite{Tong:2002rq, Harvey:2005ab, Okuyama:2005gx}.
These days the $\N=(4,4)$ GLSM for the exotic $5^2_2$-brane is also established \cite{Kimura:2013fda}.
In an $\N=(4,4)$ GLSM, an $\N=(4,4)$ vector multiplet contains an $\N=(2,2)$ vector superfield $V$ and an $\N=(2,2)$ chiral superfield $\Phi$ which takes values in the adjoint representation under gauge symmetry.
In the GLSM for the exotic $5^2_2$-brane, 
the chiral superfield $\Phi$, whose definition is $0 = \ol{D}{}_{\pm} \Phi$, 
is re-expressed by a complex superfield $C$ in such a way that
\begin{align}
\Phi \ &= \ 
\ol{D}{}_+ \ol{D}{}_- C
\, . \label{Phi2C}
\end{align}
Here we refer to $C$ as the prepotential
of the chiral superfield $\Phi$.
By using this expression, we have succeeded in dualizing a chiral superfield $\Psi$ coupled to $\Phi$ as $\Psi \Phi$ in the F-term \cite{Kimura:2013fda}.
(We consider only abelian gauge symmetry, as in \cite{Kimura:2013fda}.)
This prepotential 
carries many redundant degrees of freedom, some of which do not contribute to physics at all.
However, other
redundant component fields {\it do} play a crucial role in generating the nongeometric structure of the target space geometry of the IR sigma model.

In order to distinguish the relevant degrees of freedom from irrelevant ones in the prepotential,
we should think of the structure of the prepotential $C$ itself.
There exists a supergauge symmetry of $C$ under the definition (\ref{Phi2C}) if the supergauge transformation parameter is removed by the supercovariant derivatives $\ol{D}{}_+ \ol{D}{}_-$.
This implies that the supergauge parameter is expressed as a complex linear superfield, which is one kind of reducible superfields \cite{Gates:1984nk, Grisaru:1997ep}.
In the case of \cite{Kimura:2013fda}, however,
we cannot use the complex linear superfield itself as the supergauge parameter.
This is because the degrees of freedom of the complex linear superfield remove
the auxiliary vectorial fields which play an important role in generating the nongeometric structure.
We have to explore a suitable supergauge parameter whose degrees of freedom are less than that of the complex linear superfield.
It is possible to remove truly irrelevant degrees of freedom of the prepotential $C$, 
and we can find the relevant degrees of freedom for the sigma model for the exotic five-brane, i.e., the four complex bosonic fields and four Weyl fermions in $C$, rather than two complex bosons and two Weyl fermions in the original chiral superfield $\Phi$.
This is the main discussion of this paper. 

It is also known that other
reducible superfields, for instance, semichiral superfields ${\mathbb X}_L$ and ${\mathbb X}_R$ 
also play a central role in studying a string propagating on a generalized K\"{a}hler geometry
(see, for instance, \cite{Buscher:1987uw, Lindstrom:2004hi} and the recent work \cite{Crichigno:2015pma}).
Indeed the sigma model of the semichiral superfields has been developed in the string theory compactified on nongeometric backgrounds. 
It is interesting that the degrees of freedom of the gauge-fixed prepotential
is the same as those of the semichiral superfields.
However, we will find that the gauge-fixed prepotential 
differs from the semichiral superfields, 
and that both of them will contribute to the string theory on nongeometric backgrounds in a different manner, respectively.

%%%%%%%%%%%%%%%%%%%%%%%%%%%%%
\vspace{3mm}

The structure of this paper is as follows.
In section \ref{sect:GF},
we study a variation of the prepotential
$C$ in terms of irreducible superfields.
In order to reduce redundant degrees of freedom of $C$, we interpret this variation as a supergauge transformation. 
We investigate two gauge-fixing conditions, one of which is suitable for the $\N=(4,4)$ GLSM for the exotic five-brane \cite{Kimura:2013fda}.
We find that the gauge-fixed prepotential
is different from semichiral superfields which provide sigma models of nongeometric string backgrounds.
In section \ref{sect:dSUSY},
we discuss the supersymmetry transformations of the prepotential
$C$ and other superfields coupled to this.
This is necessary because the supersymmetry transformation breaks a gauge-fixing.
Then we have to modify the transformation of $C$ which serves as the gauge-fixing condition.
Other superfields coupled to $C$ also receive the modification of the transformation. 
We analyze the modification in the case of the $\N=(4,4)$ GLSM for the exotic five-brane.
Section \ref{sect:summary} is devoted to a summary of this work. %and discussions.
In appendix \ref{app:conventions},
we exhibit the conventions in this paper.
In appendix \ref{app:SUSYtr},
we write down the supersymmetry transformation rules of superfields in two-dimensional $\N=(2,2)$ theory.
In appendix \ref{app:GLSM},
we explicitly describe the $\N=(4,4)$ GLSM for the exotic five-brane discussed in \cite{Kimura:2013fda}.

%%%%%%%%%%%%%%%%%%%%%%%%%%%%%%%%%%%%%%%%%%%%%%%%%%%%%%%%%%%%%%%%%%%%%%%
%\newpage
\section{Gauge-fixing conditions}
\label{sect:GF}

In this section we discuss a supergauge transformation of the prepotential $C$.
The supergauge transformation should be irrelevant for the original chiral superfield $\Phi$, by definition.
In other words, the chiral superfield should be invariant under the supergauge transformation.
Originally, $C$ carries eight complex bosons and eight Weyl fermions.
Because of the supergauge symmetry, there exist a number of redundant component fields in the prepotential.
We try to reduce some of them and find an appropriate condition which generates the nongeometric structure in the IR limit of the GLSM for the exotic five-brane \cite{Kimura:2013fda}.

%%%%%%%%%%%%%%%%%%%%%%%%%%
%\newpage
\subsection{Gauge symmetry}
\label{sect:GF-SF}

The original chiral superfield $\Phi = \ol{D}{}_+ \ol{D}{}_- C$ (\ref{Phi2C}) is equivalent to
\bsubeq
\begin{gather}
\Phi \ = \ 
\ol{D}{}_+ \ol{D}{}_- \big( C + \wt{C} \big)
\, , \label{Phi2C-2}
\end{gather}
if an ambiguity $\wt{C}$ is subject to the constraint
\begin{gather}
\ol{D}{}_+ \ol{D}{}_- \wt{C} \ = \ 0
\, . \label{cond-C}
\end{gather}
\esubeq
This implies that $\wt{C}$ would be a complex linear superfield \cite{Gates:1984nk, Grisaru:1997ep} as one kind of reducible superfields\footnote{See the explicit form of the complex linear superfield $L$ in (\ref{CLS}).}.
In this work we interpret $\wt{C}$ as a supergauge transformation parameter which does not affect the original chiral superfield $\Phi$.
We can describe $\wt{C}$ in terms of $\N=(2,2)$ irreducible superfields such as \cite{Witten:1993yc}:
\begin{align*}
\text{\renewcommand{\arraystretch}{1.2}
\begin{tabular}{rll}
$X$ : & chiral & $\ol{D}{}_{\pm} X = 0$
\\
$Y$ : & twisted chiral & $\ol{D}{}_+ Y = 0 = D_- Y$
\\
$\ol{Z}$ : & anti-twisted chiral & $\ol{D}{}_- \ol{Z} = 0 = D_+ \ol{Z}$
\\
$V$ : & vector & $V^{\dagger} = V$
\end{tabular}
}
\end{align*}
Since we impose the condition (\ref{cond-C}), vector superfields cannot be a constituent of $\wt{C}$.
For the same reason, we also cannot use anti-chiral superfields.
Since the complex linear superfield carries six complex bosons and six Weyl fermions,
a generic form of $\wt{C}$ can be given as
\begin{align}
\wt{C}
\ &= \ 
X + Y + \ol{Z}
\, . \label{C=XYZ}
\end{align}
We often denote this supergauge parameter as $\wt{C} = (X,Y,\ol{Z})$.
Adding $\wt{C}$ to $C$ is interpreted as a gauge transformation.
We comment that it is not necessary that the anti-twisted chiral superfield $\ol{Z}$ is independent of the twisted chiral superfield $Y$.  
We will discuss a suitable pair of $(X,Y,\ol{Z})$ for $(\Phi, C)$ in the $\N=(4,4)$ GLSM for an exotic five-brane \cite{Kimura:2013fda}.

%%%%%%%%%%%%%%%%%%%%%%%%%%
%\newpage
\subsection{Gauge-fixing conditions}
\label{sect:GF-CF}

%Here we discuss a suitable choice of $(X,Y,\ol{Z})$. 

The explicit forms of $\Phi$ and $C$ expanded by the Grassmann coordinates $(\theta^{\pm}, \ol{\theta}{}^{\pm})$ are given as\footnote{For the convention used for superfields, see appendix \ref{app:conventions}.}
\bsubeq
\begin{align}
\Phi 
\ &= \ 
\phi
+ \I \sqrt{2} \, \theta^+ \wt{\lambda}_{+} 
+ \I \sqrt{2} \, \theta^- \wt{\lambda}_{-}
+ 2 \I \, \theta^+ \theta^- D_{\Phi}
\nn \\
\ & \ \ \ \ 
- \I \, \theta^+ \ol{\theta}{}^+ \del_+ \phi
- \I \, \theta^- \ol{\theta}{}^- \del_- \phi
+ \sqrt{2} \, \theta^+ \ol{\theta}{}^+ \theta^- \del_+ \wt{\lambda}_{-}
+ \sqrt{2} \, \theta^- \ol{\theta}{}^- \theta^+ \del_- \wt{\lambda}_{+}
\nn \\
\ & \ \ \ \
+ \theta^+ \theta^- \ol{\theta}{}^+ \ol{\theta}{}^- \del_+ \del_- \phi 
\, , \\
%%%%%%%%%%%%%%%
C \ &= \ 
\phi_{c} 
+ \I \sqrt{2} \, \theta^+ \psi_{c+} 
+ \I \sqrt{2} \, \theta^- \psi_{c-} 
+ \I \sqrt{2} \, \ol{\theta}{}^+ \chi_{c+} 
+ \I \sqrt{2} \, \ol{\theta}{}^- \chi_{c-}
\nn \\
\ & \ \ \ \ 
+ \I \, \theta^+ \theta^- F_{c} 
+ \I \, \ol{\theta}{}^+ \ol{\theta}{}^- M_{c}
+ \theta^+ \ol{\theta}{}^- G_{c} 
+ \theta^- \ol{\theta}{}^+ N_{c}
%\nn \\
%
%\ & \ \ \ \ 
+ \theta^- \ol{\theta}{}^- A_{c=}
+ \theta^+ \ol{\theta}{}^+ B_{c\+}
\nn \\
\ & \ \ \ \ 
- \sqrt{2} \, \theta^+ \theta^- \ol{\theta}{}^+ \zeta_{c+}
- \sqrt{2} \, \theta^+ \theta^- \ol{\theta}{}^- \zeta_{c-}
- \sqrt{2} \, \theta^+ \ol{\theta}{}^+ \ol{\theta}{}^- \lambda_{c+}
- \sqrt{2} \, \theta^- \ol{\theta}{}^+ \ol{\theta}{}^- \lambda_{c-}
\nn \\
\ & \ \ \ \ 
- 2 \theta^+ \theta^- \ol{\theta}{}^+ \ol{\theta}{}^- D_{c}
\, . \label{expand-C}
\end{align}
\esubeq
Here $\del_{\pm} \equiv \del_0 \pm \del_1$ are the coordinate derivatives.
The original chiral superfield $\Phi$ contains one complex scalar $\phi$, two Weyl spinors $\wt{\lambda}{}_{\pm}$ and one complex auxiliary scalar $D_{\Phi}$, while the prepotential
$C$ involves six complex scalars $(\phi_c, F_c, M_c, G_c, N_c, D_c)$, eight Weyl spinors $(\psi_{c\pm}, \chi_{c\pm}, \zeta_{c\pm}, \lambda_{c\pm})$, and two vectorial fields $(A_{c=}, B_{c\+})$ that can be regarded as light-cone coordinates of a complex vector field $W_{c,m}$.
We can read off the relations between their component fields via (\ref{Phi2C}):
\bsubeq \label{Phi2C-2}
\begin{align}
\phi \ &= \ - \I M_c
\, , \\
%%%%%%%%%%%%
D_{\Phi} \ &= \ 
- \I D_{c} + \half \del_+ A_{c=} + \half \del_- B_{c\+} + \frac{\I}{2} \del_+ \del_- \phi_{c} 
\, , \\
%%%%%%%%%%%%
\wt{\lambda}{}_{\pm}
\ &= \ 
- \I \Big( \lambda_{c\pm} \pm \del_{\pm} \chi_{c\mp} \Big)
\, , \\
%%%%%%%%%%%%
&\ls
\{ \, F_{c} \, , \ \ 
G_{c} \, , \ \ 
N_{c} \, , \ \ 
\psi_{c\pm} \, , \ \ 
\zeta_{c\pm} \, \} \, : \ \ \ 
\text{(no relations)}
\, . 
\end{align}
\esubeq
There are many redundant component fields in $C$, which should be gauged away by the equivalence between (\ref{Phi2C}) and (\ref{Phi2C-2}) under the constraint (\ref{cond-C}).
This means that we impose a gauge-fixing condition on the prepotential $C$.
Indeed this procedure is parallel to imposing the Wess-Zumino gauge on a vector superfield.
In the following discussions, we will perform two gauge-fixings. 
One is the full gauge-fixing by the supergauge parameter $\wt{C} = (X,Y,\ol{Z})$, where $Z$ is independent of $Y$.
This is equivalent to $\wt{C}$ being given by a complex linear superfield $L$, although we do not use its explicit form.
The other is a relaxed gauge-fixing given by $\wt{C} = (X, Y, -\I \ol{Y})$.
We note that the reason why the former gauge-fixing is called ``full'' is that the degrees of freedom of $C$ fixed by $\wt{C} = (X,Y,\ol{Z})$ are equal to those of the original chiral superfield $\Phi$.
In the latter case, however, the gauge-fixed degrees of freedom of $C$ are reduced by $Z = \I Y$.

%%%%%%%%%%%%%%%%%%
\subsubsection*{Full gauge-fixing by $\wt{C}=(X,Y,\ol{Z})$}

First, we consider the full gauge-fixing given by the supergauge parameter $\wt{C} = (Y,Y,\ol{Z})$. 
In this case the superfield gauge parameters $X$, $Y$ and $\ol{Z}$ are independent of each other.
By using their component fields 
(see (\ref{expand}) in appendix \ref{app:SF} for the expansions of a chiral superfield $X$ and twisted chiral superfields $Y$ and $Z$), 
we expand the transformed prepotential
$C' \equiv C + \wt{C}$ as 
\begin{align}
C' \ &= \ 
\Big( \phi_{c} + \phi_X + \sigma_Y + \ol{\sigma}{}_{Z} \Big)
+ \I \sqrt{2} \, \theta^+ \Big( \psi_{c+} + \psi_{X+} + \ol{\chi}{}_{Y+} \Big)
+ \I \sqrt{2} \, \theta^- \Big( \psi_{c-} + \psi_{X-} - \ol{\chi}{}_{Z-} \Big)
\nn \\
\ & \ \ \ \ 
+ \I \sqrt{2} \, \ol{\theta}{}^+ \Big( \chi_{c+} + \chi_{Z+} \Big)
+ \I \sqrt{2} \, \ol{\theta}{}^- \Big( \chi_{c-} - \chi_{Y-} \Big)
\nn \\
\ & \ \ \ \ 
+ \I \, \theta^+ \theta^- \Big( F_{c} + 2 F_X \Big)
+ \I \, \ol{\theta}{}^+ \ol{\theta}{}^- M_{c}
+ \theta^+ \ol{\theta}{}^- \Big( G_{c} + 2 \I \, G_Y \Big)
+ \theta^- \ol{\theta}{}^+ \Big( N_{c} - 2 \I \, \ol{G}{}_Z \Big)
\nn \\
\ & \ \ \ \ 
+ \theta^- \ol{\theta}{}^- \Big\{ A_{c=}
- \I \, \del_- \Big( 
  \phi_X
- \sigma_Y
+ \ol{\sigma}{}_{Z}
\Big)
\Big\}
+ \theta^+ \ol{\theta}{}^+ \Big\{ B_{c\+}
- \I \, \del_+ \Big(
  \phi_X
+ \sigma_Y
- \ol{\sigma}{}_{Z}
\Big)
\Big\}
\nn \\
\ & \ \ \ \ 
- \sqrt{2} \, \theta^+ \theta^- \ol{\theta}{}^+ \Big\{
  \zeta_{c+}
+ \del_+ \Big( \psi_{X-} + \ol{\chi}{}_{Z-} \Big)
\Big\}
- \sqrt{2} \, \theta^+ \theta^- \ol{\theta}{}^- \Big\{
  \zeta_{c-}
- \del_- \Big( \psi_{X+} - \ol{\chi}{}_{Y+} \Big)
\Big\}
\nn \\
\ & \ \ \ \ 
- \sqrt{2} \, \theta^+ \ol{\theta}{}^+ \ol{\theta}{}^- \Big( \lambda_{c+} + \del_+ \chi_{Y-} \Big)
- \sqrt{2} \, \theta^- \ol{\theta}{}^+ \ol{\theta}{}^- \Big( \lambda_{c-} + \del_- \chi_{Z+} \Big)
\nn \\
\ & \ \ \ \ 
- 2 \theta^+ \theta^- \ol{\theta}{}^+ \ol{\theta}{}^- \Big\{
  D_{c}
- \half \del_+ \del_- \Big( 
  \phi_X
- \sigma_Y
- \ol{\sigma}{}_Z
\Big)
\Big\}
\, . \label{C+XYZ}
\end{align}
For convenience, we also formally represent the components of $C'$ as
\begin{align}
C' \ &= \ 
\phi'_c + \I \sqrt{2} \, \theta^+ \psi'_{c+} + \ldots
\, .
\end{align}
Immediately we find the coincidence
\begin{align}
M_c' \ &= \ M_c
\, .
\end{align}
This implies that the scalar field $M_c$ is invariant under the supergauge transformation $C \to C' = C + \wt{C}$.
This is consistent with the invariance of the original chiral superfield $\Phi$ under this transformation.
The other component fields of $C$ are transformed.
Due to the expression (\ref{C+XYZ}), 
we can choose a gauge on $(F_c, G_c, N_c)$:
\bsubeq \label{FGN-1}
\begin{align}
F'_c \ &= \ F_c + 2 F_X \ \equiv \ 0
\, , \\
%%%%%%%%
G'_c \ &= \ G_c + 2 \I G_Y \ \equiv \ 0
\, , \\
%%%%%%%%
N'_c \ &= \ N_c - 2 \I \ol{G}{}_Z \ \equiv \ 0
\, .
\end{align}
\esubeq
A gauge-fixing on $(\phi_c, A_{c=}, B_{c\+}, D_c)$ is also performed easily. 
Here we introduce the following expressions:
\bsubeq
\begin{alignat}{2}
\phi'_c \ &= \ \phi_c + \wt{\phi}_c
\, , &\ls
\wt{\phi}_c \ &\equiv \ 
\phi_X + \sigma_Y + \ol{\sigma}{}_Z
\, , \\
%%%%%%%%%
A'_{c=} \ &= \ 
A_{c=} - \I \del_- \wt{A}_{c}
\, , &\ls
\wt{A}_c \ &\equiv \ 
\phi_X - \sigma_Y + \ol{\sigma}{}_Z
\, , \\
%%%%%%%%%
B'_{c\+} \ &= \ 
B_{c\+} - \I \del_+ \wt{B}_c
\, , &\ls
\wt{B}_c \ &\equiv \ 
\phi_X + \sigma_Y - \ol{\sigma}{}_Z
\, , \\
%%%%%%%%%
D'_c \ &= \ 
D_c - \half \del_+ \del_- \wt{D}_c
\, , &\ls
\wt{D}_c \ &\equiv \ 
\phi_X - \sigma_Y - \ol{\sigma}{}_Z
\, . 
\end{alignat}
\esubeq
Here we find a relation among the gauge parameters $(\phi_X, \sigma_Y, \ol{\sigma}{}_Z)$,
\begin{align}
0 \ &= \ 
\wt{\phi}{}_c - \wt{A}_c - \wt{B}_c + \wt{D}_c
\, . \label{const-XYZ}
\end{align}
This indicates that three of the four component fields $(\phi'_c, A'_{c=}, B'_{c\+}, D'_c)$ can be independently gauged away, while one of them remains non-trivial.
For instance, we can choose a gauge,
\bsubeq \label{phiABD-1}
\begin{align}
\phi'_c \ &\equiv \ 0
\, , \\
%%%%%%%%%
D'_c \ &\equiv \ 0
\, , \label{Dc=0} \\
%%%%%%%%%
{\rm Im} (A'_{c=})
\ &\equiv \ 0
\, , \\
%%%%%%%%%
{\rm Im} (B'_{c\+})
\ &\equiv \ 0
\, ,
\end{align}
\esubeq
while the real parts of the vectorial fields $(A'_{c=}, B'_{c\+})$ are non-trivial.
They, rather than their field strength, 
appear
in the scalar potential with coupling to non-dynamical scalar fields as well as dynamical ones.
Indeed, the non-dynamical field wraps a compact circle of the target space by virtue of the vectorial fields.
Here we explain this phenomenon in the GLSM for the exotic five-brane (see appendix \ref{app:GLSM} for the detailed forms). 
The non-dynamical vectorial fields $(A_{c=}, B_{c\+})$ generate a potential term such as the fifth line in (\ref{GLSM-522-b2}).
This term connects the original coordinate field $r^2$ of a compact circle and the new dualized coordinate field $y^2$ via the duality relations (\ref{r2+y2}) and (\ref{r2-y2}).
When we study supersymmetric vacua, this potential term also vanishes. 
Combining this with (\ref{r2+y2}) and (\ref{r2-y2}), this vanishing condition yields\footnote{To make the duality clear, we consider a simple case $k = 1$ of a general gauge theory (\ref{GLSM-522}).} 
\begin{align}
0 \ &= \ 
\frac{g^2}{2} \big(A_{c=} + \ol{A}{}_{c=} \big) \big(B_{c\+} + \ol{B}{}_{c\+} \big)
\nn \\
\ &= \ 
- \frac{1}{2 g^2} (\del_m r^2)^2 
+ \frac{g^2}{2} (\del_m y^2)^2 
+ \ve^{mn} (\del_m r^2) (\del_n y^2)
\, . \label{const-AcBc}
\end{align}
This implies that the dynamics of the original field $r^2$ in the GLSM moves to that of the dual field $y^2$, whilst 
the topological term still carries the original field.
Due to this relation, $r^2$ and $y^2$ seem to co-exist in the system.
However, we can integrate out $r^2$ 
in (\ref{const-AcBc})
in the low energy limit and obtain the correct nonlinear sigma model containing $y^2$. 
Furthermore, the target space metric becomes multi-valued with respect to other scalar coordinate fields \cite{Kimura:2013fda}.
On the other hand, if the real part of the non-dynamical vectorial fields $({\rm Re} A_{c=}, {\rm Re} B_{c\+})$ are absent by the gauge-fixing, we cannot obtain the correct sigma model.
Hence it turns out that the non-dynamical fields 
$(A_{c=}, B_{c\+})$
generate
the nongeometric structure on the target space geometry of the sigma model \cite{Kimura:2013fda}.
This phenomenon can be seen as an Aharonov-Bohm--like effect\footnote{While this analysis has been seen in the GLSM for the exotic five-brane, 
multi-valudness should also appear in the T-duality transformation of other defect branes of codimension two.}.
We can also introduce a gauge on the fermionic fields as
\bsubeq \label{fermions-1}
\begin{align}
\chi'_{c+} \ &= \ 
\chi_{c+} + \chi_{Z+} \ \equiv \ 0
\, , \\
%%%%%%%%
\chi'_{c-} \ &= \ 
\chi_{c-} - \chi_{Y-} \ \equiv \ 0
\, , \\
%%%%%%%%
\psi'_{c+} \ &= \ 
\psi_{c+} + (\psi_{X+} + \ol{\chi}{}_{Y+}) \ \equiv \ 0
\, , \\
%%%%%%%%
\psi'_{c-} \ &= \ 
\psi_{c-} + (\psi_{X-} - \ol{\chi}{}_{Z-}) \ \equiv \ 0
\, , \\
%%%%%%%%
\zeta'_{c+} \ &= \ 
\zeta_{c+} + \del_+ (\psi_{X-} + \ol{\chi}{}_{Z-})
\ \equiv \ 0
\, , \\
%%%%%%%%
\zeta'_{c-} \ &= \ 
\zeta_{c-} - \del_- (\psi_{X+} - \ol{\chi}{}_{Y+})
\ \equiv \ 0
\, .
\end{align}
\esubeq
The non-trivial degrees of freedom are carried only by $\lambda_{c\pm}$.
Then we find that the coincidence (\ref{Phi2C-2}) is reduced to
\bsubeq \label{Phi2C-red1}
\begin{align}
\phi \ &= \ 
- \I M'_c
\, , \\
D_{\Phi}
\ &= \ 
\half \del_+ {\rm Re} (A'_{c=}) + \half \del_- {\rm Re}(B'_{c\+})
\, , \\
\wt{\lambda}_{\pm}
\ &= \ 
- \I \lambda'_{c\pm}
\, .
\end{align}
\esubeq
Under the gauge-fixings 
(\ref{FGN-1}),
(\ref{phiABD-1}), 
and (\ref{fermions-1}) 
demonstrated above, 
we completely removed all the redundant degrees of freedom of $C$.
However, we {\it cannot} use this gauge-fixing in the $\N=(4,4)$ GLSM for the exotic five-brane \cite{Kimura:2013fda}.
This is because, 
by the reduction to 
(\ref{Dc=0}) and (\ref{Phi2C-red1}), the FI parameter $s$ coupled to $D_c$ does not contribute to the system any more (see the Lagrangian (\ref{L-PhiC-2})), while the FI parameter represents the position of a five-brane on the target space of the IR sigma model.
Simultaneously, the coupling
$(D_c - \ol{D}{}_c) r^2$ in (\ref{L-PsiC-2}), 
which plays a crucial role in generating multi-valuedness of the target space metric of the IR sigma model \cite{Kimura:2013fda}, also disappears.
Even though we introduce other gauge-fixing in which $D_c$ does not vanish,
one of the (real part of) vectorial fields $(A_{c=}, B_{c\+})$ is fixed to zero.
In this case, the Aharonov-Bohm--like effect disappears in the GLSM for the exotic five-brane, and we cannot obtain any nongeometric structure in the IR limit of the theory.
In order to obtain an appropriate gauge-fixing with non-vanishing $(D_c, {\rm Re} (A_{c=}), {\rm Re} (B_{c\+}))$, we should consider a relaxed gauge-fixing given by $\wt{C} = (X, Y, -\I \ol{Y})$.

%%%%%%%%%%%%%%%%%%
\subsubsection*{A relaxed gauge-fixing by $\wt{C} = (X,Y,-\I \ol{Y})$}

Second, we consider a relaxed gauge-fixing given by the supergauge parameter $\wt{C} = (Y,Y, -\I \ol{Y})$. 
In this case the transformed prepotential
$C' = C + \wt{C}$ is expanded as
\begin{align}
C' \ &= \
\Big(  \phi_{c} + \phi_X + \sigma_Y - \I \ol{\sigma}{}_{Y} \Big)
+ \I \sqrt{2} \, \theta^+ \Big(  \psi_{c+} + \psi_{X+} + \ol{\chi}{}_{Y+} \Big)
+ \I \sqrt{2} \, \theta^- \Big(  \psi_{c-} + \psi_{X-} + \I \ol{\chi}{}_{Y-} \Big)
\nn \\
\ & \ \ \ \ 
+ \I \sqrt{2} \, \ol{\theta}{}^+ \Big(  \chi_{c+} - \I \chi_{Y+} \Big)
+ \I \sqrt{2} \, \ol{\theta}{}^- \Big(  \chi_{c-} - \chi_{Y-} \Big)
\nn \\
\ & \ \ \ \ 
+ \I \, \theta^+ \theta^- \Big(  F_{c} + 2 F_X \Big)
+ \I \, \ol{\theta}{}^+ \ol{\theta}{}^-  M_{c}
+ \theta^+ \ol{\theta}{}^- \Big(  G_{c} + 2 \I \, G_Y \Big)
+ \theta^- \ol{\theta}{}^+ \Big(  N_{c} - 2 \, \ol{G}{}_Y \Big)
\nn \\
\ & \ \ \ \ 
+ \theta^- \ol{\theta}{}^- \Big\{  A_{c=}
- \I \, \del_- \Big( 
  \phi_X
- \sigma_Y
- \I \ol{\sigma}{}_{Y}
\Big)
\Big\}
+ \theta^+ \ol{\theta}{}^+ \Big\{  B_{c\+}
- \I \, \del_+ \Big(
  \phi_X
+ \sigma_Y
+ \I \ol{\sigma}{}_{Y}
\Big)
\Big\}
\nn \\
\ & \ \ \ \ 
- \sqrt{2} \, \theta^+ \theta^- \ol{\theta}{}^+ \Big\{
   \zeta_{c+}
+ \del_+ \Big( \psi_{X-} - \I \ol{\chi}{}_{Y-} \Big)
\Big\}
- \sqrt{2} \, \theta^+ \theta^- \ol{\theta}{}^- \Big\{
   \zeta_{c-}
- \del_- \Big( \psi_{X+} - \ol{\chi}{}_{Y+} \Big)
\Big\}
\nn \\
\ & \ \ \ \ 
- \sqrt{2} \, \theta^+ \ol{\theta}{}^+ \ol{\theta}{}^- \Big( 
 \lambda_{c+} 
+ \del_+ \chi_{Y-} 
\Big)
- \sqrt{2} \, \theta^- \ol{\theta}{}^+ \ol{\theta}{}^- \Big( 
 \lambda_{c-} 
- \I \del_- \chi_{Y+} 
\Big)
\nn \\
\ & \ \ \ \ 
- 2 \theta^+ \theta^- \ol{\theta}{}^+ \ol{\theta}{}^- \Big\{
  D_{c}
- \half \del_+ \del_- \Big( 
  \phi_X
- \sigma_Y
+ \I \ol{\sigma}{}_Y
\Big)
\Big\}
\, . \label{C+XYY}
\end{align}
Again the scalar field $M_c$ is invariant.
Let us focus on the transformation rules of $(F_c, G_c, N_c)$.
In the previous case all of them can be gauged away by three independent gauge parameters $(F_X, G_Y, \ol{G}{}_Z)$. 
However, in the present case, both of $(G_c, N_c)$ cannot be simultaneously gauged away, while $F_c$ can be removed. 
For instance, we take the following gauge,
\bsubeq \label{FGN-2}
\begin{align}
F'_c \ &= \ F_c + 2 F_X \ \equiv \ 0
\, , \\
%%%%%%%%
N'_c \ &= \ N_c - 2 \ol{G}{}_Y \ \equiv \ 0
\, , \\
%%%%%%%%
G'_c \ &= \ G_c + 2 \I G_Y \ \neq \ 0
\, .
\end{align}
\esubeq
We study a gauge-fixing on $(\phi_c, A_{c=}, B_{c\+}, D_c)$.
As in the previous discussion, we consider their gauge transformations:
\bsubeq
\begin{alignat}{2}
\phi'_c \ &= \ \phi_c + \wt{\phi}_c
\, , &\ls
\wt{\phi}_c \ &\equiv \ 
\phi_X + \sigma_Y - \I \ol{\sigma}{}_Y
\, , \\
%%%%%%%%%
A'_{c=} \ &= \ 
A_{c=} - \I \del_- \wt{A}_{c}
\, , &\ls
\wt{A}_c \ &\equiv \ 
\phi_X - \sigma_Y - \I \ol{\sigma}{}_Y
\, , \\
%%%%%%%%%
B'_{c\+} \ &= \ 
B_{c\+} - \I \del_+ \wt{B}_c
\, , &\ls
\wt{B}_c \ &\equiv \ 
\phi_X + \sigma_Y + \I \ol{\sigma}{}_Y
\, , \\
%%%%%%%%%
D'_c \ &= \ 
D_c - \half \del_+ \del_- \wt{D}_c
\, , &\ls
\wt{D}_c \ &\equiv \ 
\phi_X - \sigma_Y + \I \ol{\sigma}{}_Y
\, . 
\end{alignat}
\esubeq 
Again there exists a relation (\ref{const-XYZ}).
Counting the degrees of freedom, 
we can set two of the fields $(\phi'_c, A'_{c=}, B'_{c\+}, D'_c)$ to zero by the parameters $(\phi_X, \sigma_Y)$.
Thus we adopt the following gauge-fixing:
\bsubeq \label{phiABD-2}
\begin{align}
\phi'_c \ &\equiv \ 0
\, , \\
%%%%%%%%%%
{\rm Im} (A'_{c=}) \ &\equiv \ 0
\, , \\
%%%%%%%%%%
{\rm Im} (B'_{c\+}) \ &\equiv \ 0
\, , \\
%%%%%%%%%%
{\rm Re} (A'_{c=})
\ &= \
{\rm Re} (A_{c=})
+ {\rm Re} (\wt{\phi}_c)
+ {\rm Im} (\wt{\phi}_c)
- {\rm Im} (\wt{B}_c)
\ \neq \ 0
\, , \\
%%%%%%%%%%
{\rm Re} (B'_{c\+})
\ &= \ 
{\rm Re} (B_{c\+})
+ {\rm Re} (\wt{\phi}_c)
- {\rm Im} (\wt{\phi}_c)
- {\rm Im} (\wt{A}_c)
+ 2 {\rm Im} (\wt{B}_c)
\ \neq \ 0
\, , \\
%%%%%%%%%%
D'_c \ &= \ 
D_c 
- {\rm Im} (\wt{\phi}_c)
+ {\rm Im} (\wt{A}_c)
+ {\rm Im} (\wt{B}_c)
\ \neq \ 0
\, .
\end{align}
\esubeq
We also study a gauge-fixing on the fermions.
Here we introduce the following gauge:
\bsubeq \label{fermions-2}
\begin{align}
\zeta'_{c+} + \del_+ \psi'_{c-}
\ &= \ 
\zeta_{c+} + \del_+ \psi_{c-} + 2 \del_+ \psi_{X-}
\ \equiv \ 0
\, , \\
%%%%%%%%%%%
\zeta'_{c-} - \del_- \psi'_{c+}
\ &= \ 
\zeta_{c-} - \del_- \psi_{c+} - 2 \del_- \psi_{X+}
\ \equiv \ 0
\, , \\
%%%%%%%%%%%
\lambda'_{c+} - \del_+ \chi'_{c-}
\ &= \ 
\lambda_{c+} - \del_+ \chi_{c-} + 2 \del_+ \chi_{Y-}
\ \equiv \ 0
\, , \\
%%%%%%%%%%%
\lambda'_{c-} + \del_- \chi'_{c+}
\ &= \ 
\lambda_{c-} + \del_- \chi_{c+} - 2 \I \del_- \chi_{Y+}
\ \equiv \ 0
\, , \\
%%%%%%%%%%%
\zeta'_{c+} - \del_+ \psi'_{c-}
\ &= \ 
\zeta_{c+} - \del_+ \psi_{c-} - 2 \I \del_+ \ol{\chi}{}_{Y-}
\ \neq \ 0
\, , \\
%%%%%%%%%%%
\zeta'_{c-} + \del_- \psi'_{c+}
\ &= \ 
\zeta_{c-} + \del_- \psi_{c+} + 2 \del_- \ol{\chi}{}_{Y+}
\ \neq \ 0
\, , \\
%%%%%%%%%%%
\lambda'_{c+} + \del_+ \chi'_{c-}
\ &= \ 
\lambda_{c+} + \del_+ \chi_{c-}
\ \neq \ 0
\, , \\
%%%%%%%%%%%
\lambda'_{c-} - \del_- \chi'_{c+}
\ &= \ 
\lambda_{c-} - \del_- \chi_{c+} 
\ \neq \ 0
\, .
\end{align}
\esubeq
This gauge-fixing removes any derivative interaction terms of 
the fermionic fields.
Furthermore, we will obtain appropriate supersymmetry transformations consistent with the gauge-fixing.
We will discuss these issues in the next section.
Due to the gauge-fixings 
(\ref{FGN-2}),
(\ref{phiABD-2})
and 
(\ref{fermions-2})
 by the supergauge parameter $\wt{C} = (X, Y, -\I \ol{Y})$, the component fields of the original chiral superfield $\Phi$ are expressed as
\bsubeq 
\begin{align}
\phi \ &= \ 
- \I M'_c
\, , \\
%%%
D_{\Phi} \ &= \ 
- \I D'_c 
+ \half \del_+ {\rm Re} (A'_{c=})
+ \half \del_- {\rm Re} (B'_{c\+})
\, , \\
%%%
\wt{\lambda}{}_{\pm}
\ &= \ 
- \I \Big( \lambda'_{c\pm} \pm \del_{\pm} \chi'_{c\mp} \Big)
\ = \ 
- 2 \I \lambda'_{c\pm}
\ = \ 
\mp 2 \I \del_{\pm} \chi'_{c\mp}
\, .
\end{align}
\esubeq
Since all $(D'_c, {\rm Re}(A'_{c=}), {\rm Re}(B'_{c\+}))$ are non-trivial, we can obtain suitable gauge-fixed Lagrangians of (\ref{L-PhiC-2}) and (\ref{L-PsiC-2}) for the sigma model with nongeometric structure \cite{Kimura:2013fda}.

Now we count the remaining component fields of the gauge-fixed prepotential $C'$.
There are two complex scalars $(M'_c, D'_c)$, two real vectorial fields $({\rm Re} (A'_{c=}), {\rm Re}(B'_{c\+}))$, 
and four Weyl fermions $(\zeta'_{c\pm} \mp \del_{\pm} \psi'_{c\mp}, \lambda'_{c\pm} \pm \del_{\pm} \chi'_{c\mp})$.
The number of degrees of freedom is still twice as many as that of the original chiral superfield $\Phi$, while the redundancy plays a significant role in the sigma model for the exotic five-brane \cite{Kimura:2013fda}.
We notice that the gauge-fixed prepotential
$C'$ is neither a left semichiral superfield ${\mathbb X}_L$ nor a right semichiral superfield ${\mathbb X}_R$ \cite{Buscher:1987uw, Lindstrom:2004hi, Crichigno:2015pma}, whose definitions are $0 = \ol{D}{}_+ {\mathbb X}_L$ and $0 = \ol{D}{}_- {\mathbb X}_R$, respectively.
The scalar components of the semichiral superfields behave as coordinates of the target space geometry which is a generalized K\"{a}hler geometry, or more general, a nongeometric background.
We insist that the gauge-fixed prepotential
$C'$ also contributes to the sigma model whose target space is a nongeometric background, though $C'$ belongs to an $\N=(4,4)$ vector multiplet and does not represent any coordinates of the target space.

%%%%%%%%%%%%%%%%%%%%%%%%%%%%%%%%%%%%%%%%%%%%%%%%%%%%%%%%%%%%%%%%%%%%%%%
%\newpage
\section{Modification of supersymmetry transformations}
\label{sect:dSUSY}

When we impose a gauge-fixing condition on the prepotential $C$,
the supersymmetry transformation of its component fields is reduced.
However, the supersymmetry transformation breaks the gauge-fixing condition.
In order to restore the gauge-fixing condition even after the supersymmetry transformation,
we have to modify the supersymmetry transformation rule by adding appropriate terms.
This situation is completely parallel to the Wess-Zumino gauge-fixing and modification of the supersymmetry transformation under the Wess-Zumino gauge.

%%%%%%%%%%%%%%%%%%%%%%%
\subsection{Supersymmetry transformation under the gauge-fixing condition}

Under the gauge-fixing condition 
(\ref{FGN-2}),
(\ref{phiABD-2})
and 
(\ref{fermions-2}),
the supersymmetry transformation of the component fields of the gauge-fixed 
prepotential $C'$ is reduced to\footnote{The supersymmetry transformation of the prepotential
without any gauge-fixing conditions is listed in appendix \ref{SUSY-C-BGF}.} 
\bsubeq \label{SUSY-C-XYY}
\begin{align}
\delta \phi'_c
\ &= \ 
  \sqrt{2} \, \ve_- \psi'_{c+} 
- \sqrt{2} \, \ve_+ \psi'_{c-} 
- \sqrt{2} \, \ol{\ve}{}_- \chi'_{c+} 
+ \sqrt{2} \, \ol{\ve}{}_+ \chi'_{c-}
%\vphantom{\frac{1}{\sqrt{2}}}
\, , \\
%%%%%%%%%%%%%%%%%%%
\delta \psi'_{c+}
\ &= \ 
  \frac{1}{\sqrt{2}} \, \ol{\ve}{}_- {\rm Re} (B'_{c\+})
- \frac{1}{\sqrt{2}} \, \ol{\ve}{}_+ G'_{c} 
\, , \\
%%%%%%%%%%%%%%%%%%%
\delta \psi'_{c-}
\ &= \
- \frac{1}{\sqrt{2}} \, \ol{\ve}{}_+ {\rm Re} (A'_{c=})
\, , \\
%%%%%%%%%%%%%%%%%%%
\delta \chi'_{c+}
\ &= \
  \frac{1}{\sqrt{2}} \, \ve_- {\rm Re} (B'_{c\+})
- \frac{\I}{\sqrt{2}} \, \ol{\ve}{}_+ M'_{c}
\, , \\
%%%%%%%%%%%%%%%%%%%
\delta \chi'_{c-}
\ &= \
- \frac{1}{\sqrt{2}} \, \ve_+ {\rm Re} (A'_{c=})
+ \frac{1}{\sqrt{2}} \, \ve_- G'_{c} 
- \frac{\I}{\sqrt{2}} \, \ol{\ve}{}_- M'_{c}
\, , \\
%%%%%%%%%%%%%%%%%%%
\delta F'_{c}
\ &= \ 
0
\, , \\
%%%%%%%%%%%%%%%%%%%
\delta M'_{c}
\ &= \ 
  \sqrt{2} \, \ve_- \Big(
  \lambda'_{c+}
+ \del_+ \chi'_{c-}
\Big)
- \sqrt{2} \, \ve_+ \Big(
  \lambda'_{c-}
- \del_- \chi'_{c+}
\Big)
%\vphantom{\frac{1}{\sqrt{2}}}
\, , \\
%%%%%%%%%%%%%%%%%%%
\delta G'_{c}
\ &= \ 
2 \sqrt{2} \, \I \, \ve_+ \del_- \psi'_{c+} 
%\vphantom{\frac{1}{\sqrt{2}}}
\, , \\
%%%%%%%%%%%%%%%%%%%
\delta N'_{c}
\ &= \ 
- 2 \sqrt{2} \, \I \, \ve_- \del_+ \psi'_{c-} 
%\vphantom{\frac{1}{\sqrt{2}}}
\, , \\
%%%%%%%%%%%%%%%%%%%
\delta A'_{c=}
\ &= \ 
+ \I \sqrt{2} \, \ve_+ \del_- \psi'_{c-} 
+ \I \sqrt{2} \, \ve_- \del_- \psi'_{c+}
+ \I \sqrt{2} \, \ol{\ve}{}_+ \del_- \chi'_{c-}
- \I \sqrt{2} \, \ol{\ve}{}_- \del_- \chi'_{c+}
%\vphantom{\frac{1}{\sqrt{2}}}
\, , \\
%%%%%%%%%%%%%%%%%%%
\delta B'_{c\+}
\ &= \ 
- \I \sqrt{2} \, \ve_- \del_+ \psi'_{c+} 
- \I \sqrt{2} \, \ve_+ \del_+ \psi'_{c-}
- \I \sqrt{2} \, \ol{\ve}{}_- \del_+ \chi'_{c+}
+ \I \sqrt{2} \, \ol{\ve}{}_+ \del_+ \chi'_{c-}
%\vphantom{\frac{1}{\sqrt{2}}}
\, , \\
%%%%%%%%%%%%%%%%%%%
\delta \zeta'_{c+}
\ &= \ 
- \frac{1}{\sqrt{2}} \, \ol{\ve}{}_+ \Big\{
  \del_- {\rm Re} (B'_{c\+})
+ 2 \I D'_{c}
\Big\}
\, , \\
%%%%%%%%%%%%%%%%%%%
\delta \zeta'_{c-}
\ &= \ 
- \frac{1}{\sqrt{2}} \, \ol{\ve}{}_- \Big\{
  \del_+ {\rm Re} (A'_{c=})
+ 2 \I D'_{c}
\Big\}
- \frac{1}{\sqrt{2}} \, \ol{\ve}{}_+ \del_- G'_{c} 
\, , \\
%%%%%%%%%%%%%%%%%%%
\delta \lambda'_{c+}
\ &= \ 
- \frac{1}{\sqrt{2}} \, \ve_+ \Big\{
  \del_- {\rm Re} (B'_{c\+})
- 2 \I D'_{c}
\Big\}
- \frac{1}{\sqrt{2}} \, \ve_- \del_+ G'_{c} 
- \frac{\I}{\sqrt{2}} \, \ol{\ve}{}_- \del_+ M'_{c}
\, , \\
%%%%%%%%%%%%%%%%%%%
\delta \lambda'_{c-}
\ &= \ 
- \frac{1}{\sqrt{2}} \, \ve_- \Big\{
  \del_+ {\rm Re} (A'_{c=})
- 2 \I D'_{c}
\Big\}
+ \frac{\I}{\sqrt{2}} \, \ol{\ve}{}_+ \del_- M'_{c}
\, , \\
%%%%%%%%%%%%%%%%%%%
\delta D'_{c}
\ &= \ 
  \frac{1}{\sqrt{2}} \, \ve_- \del_+ \del_- \psi'_{c+}
- \frac{1}{\sqrt{2}} \, \ve_+ \del_+ \del_- \psi'_{c-}
+ \frac{1}{\sqrt{2}} \, \ol{\ve}{}_- \del_+ \del_- \chi'_{c+}
- \frac{1}{\sqrt{2}} \, \ol{\ve}{}_+ \del_+ \del_- \chi'_{c-}
\, .
\end{align}
\esubeq
Since, for instance, the variation $\delta \phi'_c$ does not vanish, the supersymmetry transformation (\ref{SUSY-C-XYY}) breaks the gauge-fixing condition.
Even in this variation, however, we should keep in mind that $\delta F'_c$ still vanishes.
This implies that the transformed complex scalar $F'_c$ still satisfies the gauge-fixing condition (\ref{FGN-2}).
This will play a key role in the modification of the supersymmetry transformation.

%%%%%%%%%%%%%%%%%%%%%%%
\subsection{Modification of supersymmetry transformation of $C$}

In order that the supersymmetry transformation of the gauge-fixed prepotential
$C'$ is consistent with the gauge-fixings 
(\ref{FGN-2}),
(\ref{phiABD-2}) and
(\ref{fermions-2}),
we add the following variation to the original supersymmetry transformation $\delta C'$:
\begin{align}
\delta' C' \ &\equiv \
X' + Y' - \I \ol{Y}{}'
\, . \label{adSUSY-C-XYY}
\end{align}
Since we have chosen the supergauge parameter $\wt{C} = X + Y - \I \ol{Y}$ rather than a complex linear superfield, we should also describe the modification of the supersymmetry by a sum of irreducible superfields.
Here $X'$ and $Y'$ are a chiral superfield and a twisted chiral superfield, respectively.
Their component fields are described in the same way as in (\ref{expand}) with the prime symbol.
$\delta' C'$ represents an additional supersymmetry variation involving the supersymmetry parameters $(\ve_{\pm}, \ol{\ve}{}_{\pm})$.
Then we impose that the total variation 
\begin{align}
\wh{\delta} C' \ &\equiv \
\delta C' + \delta' C'
\, \label{totalSUSY-C}
\end{align}
should be consistent with the gauge-fixing condition 
(\ref{FGN-2}),
(\ref{phiABD-2}) and
(\ref{fermions-2}).
We notice that the additional variation $\delta' C'$ must satisfy the projection (\ref{cond-C}),
and that the pair $(X', Y')$ is different from the supergauge parameter $\wt{C} = (X,Y, -\I \ol{Y})$ in the previous section.

We expand the total variation (\ref{totalSUSY-C}) in terms of the component fields of $(C', X', Y')$. 
This is quite similar to (\ref{C+XYY}):
\begin{align}
\wh{\delta} C' 
\ &= \ 
\Big( \delta \phi'_{c} + \phi'_X + \sigma'_Y - \I \ol{\sigma}{}'_{Y} \Big)
+ \I \sqrt{2} \, \theta^+ \Big( \delta \psi'_{c+} + \psi'_{X+} + \ol{\chi}{}'_{Y+} \Big)
+ \I \sqrt{2} \, \theta^- \Big( \delta \psi'_{c-} + \psi'_{X-} + \I \ol{\chi}{}'_{Y-} \Big)
\nn \\
\ & \ \ \ \ 
+ \I \sqrt{2} \, \ol{\theta}{}^+ \Big( \delta \chi'_{c+} - \I \chi'_{Y+} \Big)
+ \I \sqrt{2} \, \ol{\theta}{}^- \Big( \delta \chi'_{c-} - \chi'_{Y-} \Big)
\nn \\
\ & \ \ \ \ 
+ 2 \I \, \theta^+ \theta^- F'_X 
+ \I \, \ol{\theta}{}^+ \ol{\theta}{}^- \delta M'_{c}
+ \theta^+ \ol{\theta}{}^- \Big( \delta G'_{c} + 2 \I \, G'_Y \Big)
+ \theta^- \ol{\theta}{}^+ \Big( \delta N'_{c} - 2 \, \ol{G}{}'_Y \Big)
\nn \\
\ & \ \ \ \ 
+ \theta^- \ol{\theta}{}^- \Big\{ \delta A'_{c=}
- \I \, \del_- \Big( 
  \phi'_X
- \sigma'_Y
- \I \ol{\sigma}{}'_{Y}
\Big)
\Big\}
+ \theta^+ \ol{\theta}{}^+ \Big\{ \delta B'_{c\+}
- \I \, \del_+ \Big(
  \phi'_X
+ \sigma'_Y
+ \I \ol{\sigma}{}'_{Y}
\Big)
\Big\}
\nn \\
\ & \ \ \ \ 
- \sqrt{2} \, \theta^+ \theta^- \ol{\theta}{}^+ \Big\{
  \delta \zeta'_{c+}
+ \del_+ \Big( \psi'_{X-} - \I \ol{\chi}{}'_{Y-} \Big)
\Big\}
- \sqrt{2} \, \theta^+ \theta^- \ol{\theta}{}^- \Big\{
  \delta \zeta'_{c-}
- \del_- \Big( \psi'_{X+} - \ol{\chi}{}'_{Y+} \Big)
\Big\}
\nn \\
\ & \ \ \ \ 
- \sqrt{2} \, \theta^+ \ol{\theta}{}^+ \ol{\theta}{}^- \Big( 
\delta \lambda'_{c+} 
+ \del_+ \chi'_{Y-} 
\Big)
- \sqrt{2} \, \theta^- \ol{\theta}{}^+ \ol{\theta}{}^- \Big( 
\delta \lambda'_{c-} 
- \I \del_- \chi'_{Y+} 
\Big)
\nn \\
\ & \ \ \ \ 
- 2 \theta^+ \theta^- \ol{\theta}{}^+ \ol{\theta}{}^- \Big\{
  \delta D'_{c}
- \half \del_+ \del_- \Big( 
  \phi'_X
- \sigma'_Y
+ \I \ol{\sigma}{}'_Y
\Big)
\Big\}
\, . \label{dh-C'}
\end{align}
The gauge-fixing condition 
(\ref{FGN-2}),
(\ref{phiABD-2}) and
(\ref{fermions-2}) 
leads to the following constraints:
\bsubeq \label{0-cond-1'}
\begin{align}
0 \ \equiv \ \wh{\delta} \phi'_c
\ &= \ 
  \sqrt{2} \, \ve_- \psi'_{c+} 
- \sqrt{2} \, \ve_+ \psi'_{c-} 
- \sqrt{2} \, \ol{\ve}{}_- \chi'_{c+} 
+ \sqrt{2} \, \ol{\ve}{}_+ \chi'_{c-}
\nn \\
%%%
\ & \ \ \ \ 
+ \phi'_X + \sigma'_Y - \I \ol{\sigma}{}'_{Y} 
\, , \\
%%%%%%%%%%%%%%%%%%%%%%%%%%%
0 \ \equiv \ \wh{\delta} F'_c
\ &= \ 
2 F'_X
\, , \\
%%%%%%%%%%%%%%%%%%%%%%%%%%%
0 \ \equiv \ \wh{\delta} N'_c
\ &= \ 
- 2 \sqrt{2} \, \I \, \ve_- \del_+ \psi'_{c-} 
- 2 \ol{G}{}'_Y
\, , \\
%%%%%%%%%%%%%%%%%%%%%%%%%%%
0 \ \equiv \ 
\wh{\delta} \, {\rm Im} (A'_{c=})
\ &= \ 
  \frac{1}{\sqrt{2}} \, \ve_+ \del_- \Big( \psi'_{c-} - \ol{\chi}{}'_{c-} \Big)
+ \frac{1}{\sqrt{2}} \, \ve_- \del_- \Big( \psi'_{c+} + \ol{\chi}{}'_{c+} \Big)
\nn \\
%%%
\ & \ \ \ \ 
- \frac{1}{\sqrt{2}} \, \ol{\ve}{}_+ \del_- \Big( \ol{\psi}{}'_{c-} - \chi'_{c-} \Big)
- \frac{1}{\sqrt{2}} \, \ol{\ve}{}_- \del_- \Big( \ol{\psi}{}'_{c+} + \chi'_{c+} \Big)
\nn \\
%%%
\ & \ \ \ \
- \half \del_- \Big( 
  \phi'_X
- \sigma'_Y
- \I \ol{\sigma}{}'_{Y}
\Big)
- \half \del_- \Big( 
  \ol{\phi}{}'_X
- \ol{\sigma}{}'_Y
+ \I \sigma'_{Y}
\Big)
\, , \\
%%%%%%%%%%%%%%%%%%%%%%%%%%%
0 \ \equiv \ 
\wh{\delta} \, {\rm Im} (B'_{c\+})
\ &= \ 
- \frac{1}{\sqrt{2}} \, \ve_+ \del_+ \Big( \psi'_{c-} + \ol{\chi}{}'_{c-} \Big)
- \frac{1}{\sqrt{2}} \, \ve_- \del_+ \Big( \psi'_{c+} - \ol{\chi}{}'_{c+} \Big)
\nn \\
%%%
\ & \ \ \ \ 
+ \frac{1}{\sqrt{2}} \, \ol{\ve}{}_+ \del_+ \Big( \ol{\psi}{}'_{c-} + \chi'_{c-} \Big)
+ \frac{1}{\sqrt{2}} \, \ol{\ve}{}_- \del_+ \Big( \ol{\psi}{}'_{c+} - \chi'_{c+} \Big)
\nn \\
%%%
\ & \ \ \ \ 
- \half \del_+ \Big(
  \phi'_X
+ \sigma'_Y
+ \I \ol{\sigma}{}'_{Y}
\Big)
- \half \del_+ \Big(
  \ol{\phi}{}'_X
+ \ol{\sigma}{}'_Y
- \I \sigma'_{Y}
\Big)
\, , \\
%%%%%%%%%%%%%%%%%%%%%%%%%%%
0 \ \equiv \ \wh{\delta} (\zeta'_{c+} + \del_+ \psi'_{c-})
\ &= \ 
- \frac{1}{\sqrt{2}} \, \ol{\ve}{}_+ \Big\{
  \del_+ {\rm Re} (A'_{c=})
+ \del_- {\rm Re} (B'_{c\+})
+ 2 \I D'_{c}
\Big\}
+ 2 \del_+ \psi'_{X-}
\, , \\
%%%%%%%%%%%%%%%%%%%%%%%%%%%
0 \ \equiv \ \wh{\delta} (\zeta'_{c-} - \del_- \psi'_{c+})
\ &= \ 
- \frac{1}{\sqrt{2}} \, \ol{\ve}{}_- \Big\{
  \del_+ {\rm Re} (A'_{c=})
+ \del_- {\rm Re} (B'_{c\+})
+ 2 \I D'_{c}
\Big\}
- 2 \del_- \psi'_{X+}
\, , \\
%%%%%%%%%%%%%%%%%%%%%%%%%%%
0 \ \equiv \ \wh{\delta} (\lambda'_{c+} - \del_+ \chi'_{c-})
\ &= \ 
- \frac{1}{\sqrt{2}} \, \ve_+ \Big\{
- \del_+ {\rm Re} (A'_{c=})
+ \del_- {\rm Re} (B'_{c\+})
- 2 \I D'_{c}
\Big\}
+ 2 \del_+ \chi'_{Y-}
\, , \\
%%%%%%%%%%%%%%%%%%%%%%%%%%%
0 \ \equiv \ \wh{\delta} (\lambda'_{c-} + \del_- \chi'_{c+})
\ &= \ 
- \frac{1}{\sqrt{2}} \, \ve_- \Big\{
  \del_+ {\rm Re} (A'_{c=})
- \del_- {\rm Re} (B'_{c\+})
- 2 \I D'_{c}
\Big\}
- 2 \I \del_- \chi'_{Y+}
\, .
\end{align}
\esubeq
Solving the above equations (\ref{0-cond-1'}),
we can write down the explicit forms of the component fields of $(X', Y')$, 
though it is not so important in this work.
Instead, we focus on the vanishing complex scalar $F'_X$.
This is a strong statement. 
The reason is as follows.
The prepotential
$C$ couples to the twisted chiral superfield $\Xi$ (see appendix \ref{app:GLSM}).
If $F'_X$ were not zero, we could not find any consistent modifications of the supersymmetry transformation of $\Xi$.
This will be investigated in the next subsection.

%%%%%%%%%%%%%%%%%%%%%%%
\subsection{Modification of supersymmetry transformations of $\Xi$ and $\Psi$}

Once we take a gauge-fixing condition on the prepotential $C$,
its supersymmetry transformation breaks the gauge-fixing condition.
In the last subsection,
we modified the supersymmetry transformation of $C$ in order to recover the gauge-fixing condition.
In this subsection we investigate the modification of the supersymmetry transformation of other superfields coupled to $C$.

Here we focus only on the $\N=(4,4)$ GLSM for the exotic five-brane \cite{Kimura:2013fda}.
The Lagrangian in the superfield formalism and its expansion by the component fields can be seen in appendix \ref{app:GLSM}.
We extract the terms in which the prepotential
$C$ couples to other superfields,
\bsubeq \label{XiPsiC}
\begin{align}
\Xi + \ol{\Xi} - \sqrt{2} \, (C + \ol{C})
\ &\in \ 
\Scr{L}_{\Xi}
\, , \label{XiC} \\
%%%%%%%%%%%%%%
(\Psi - \ol{\Psi}) (C - \ol{C})
\ &\in \ 
\Scr{L}_{\Psi C}
\, , \label{PsiC}
\end{align}
\esubeq
They are manifestly invariant under the original supersymmetry transformations unless a gauge-fixing condition is imposed (see appendix \ref{app:SUSYtr}).
Now we impose the gauge-fixing condition 
(\ref{FGN-2}),
(\ref{phiABD-2}) and
(\ref{fermions-2}) by the supergauge parameter $\wt{C} = (X,Y,-\I \ol{Y})$.
Then $C$ in (\ref{XiPsiC}) is replaced with the gauge-fixed prepotential $C'$.
Even under the gauge-fixing condition, the terms (\ref{XiPsiC}) should be invariant under the supersymmetry transformations.
This implies that the (real part of) twisted chiral superfield $\Xi$ and the (imaginary part of) chiral superfield $\Psi$ should receive the modification of the supersymmetry transformation (\ref{adSUSY-C-XYY}).
Because of this, we have to discuss the additional variations of $\Xi$ and $\Psi$.
Since the additional variation $\delta' C'$ should be canceled by the additional variations $(\delta' \Xi, \delta' \Psi)$, 
the following equations are imposed:
\bsubeq \label{SUSY'-XiPsiC}
\begin{align}
0 \ &= \ 
\big( \delta' \Xi + \delta' \ol{\Xi} \big) 
- \sqrt{2} \big( \delta' C' + \delta' \ol{C}{}' \big)
\, , \label{SUSY'-XiC} \\
%%%%%%%%%%%%%%
0 \ &= \ 
\big( \delta' \Psi - \delta' \ol{\Psi} \big) (C' - \ol{C}{}')
+ (\Psi - \ol{\Psi}) \big( \delta' C' - \delta' \ol{C}{}' \big)
\, . \label{SUSY'-PsiC}
\end{align}
\esubeq
We analyze them separately.

We have a comment before the analyses.
The Lagrangian $\Scr{L}_{\Phi}$ (\ref{L-PhiC-2}) is expressed in terms of the component fields of the prepotential $C$.
Then we should think of the invariance of this Lagrangian under the modification of supersymmetry transformations.
However, the original chiral superfield $\Phi$ is not sensitive under the modification (\ref{adSUSY-C-XYY}) because $\delta' C'$ is projected out by the condition (\ref{cond-C}).
Then we conclude that we do not have to manage the Lagrangian $\Scr{L}_{\Phi}$ carefully.

%%%%%%%%%%%%%%%%%%%%%%%%%%%%%%%
\subsubsection*{Additional variation $\delta' \Xi$}

The additional variation
$\delta' C'$ deforms the supersymmetry transformation of the real part of $\Xi$, which should be controlled by the equation (\ref{SUSY'-XiC}).
Since this is a linear equation, it is easy to obtain the form of $\delta' \Xi$ in terms of the component fields:
\bsubeq \label{C2Xi}
\begin{align}
\delta' y^1 
\ &= \ 
\big( \phi'_X + \sigma'_Y - \I \ol{\sigma}{}'_Y \big)
+ \big( \ol{\phi}{}'_X + \ol{\sigma}{}'_Y + \I \sigma'_Y \big)
\, , \label{SUSY-y1-2} \\
%%%%%%%%%%%%%%%%%%%%%%
\delta' \ol{\xi}{}_+
\ &= \ 
\sqrt{2} \, \Big(
\psi'_{X+} + \ol{\chi}{}'_{Y+} + \I \ol{\chi}{}'_{Y+}
\Big)
\, , \label{SUSY-xi+-2} \\
%%%%%%%%%%%%%%%%%%%%%%
\delta' \xi_-
\ &= \ 
- \sqrt{2} \, \Big(
  \ol{\psi}{}'_{X-}
- \chi'_{Y-}
- \I \chi'_{Y-}
\Big)
\, , \label{SUSY-xi--2} \\
%%%%%%%%%%%%%%%%%%%%%%
\del_+ \delta' y^2 
\ &= \ 
- \I \del_+ \Big\{
\big( \phi'_X + \sigma'_Y + \I \ol{\sigma}{}'_Y \big)
- \big( \ol{\phi}{}'_X + \ol{\sigma}{}'_Y - \I \sigma'_Y \big)
\Big\}
\, , \label{SUSY-d+y2-2} \\
%%%%%%%%%%%%%%%%%%%%%%
- \del_- \delta' y^2 
\ &= \ 
- \I \del_- \Big\{
\big( \phi'_X - \sigma'_Y - \I \ol{\sigma}{}'_Y \big)
- \big( \ol{\phi}{}'_X - \ol{\sigma}{}'_Y + \I \sigma'_Y \big)
\Big\}
\, , \label{SUSY-d-y2-2} \\
%%%%%%%%%%%%%%%%%%%%%%
\delta' G_{\Xi}
\ &= \ 
(1 + \I) \sqrt{2} \, G'_Y
\, . \label{SUSY-G-2}
\end{align}
\esubeq
We can also extract equations for the derivatives of $y^1$ and $\xi_{\pm}$ from (\ref{SUSY'-XiC}), while they do not satisfy the equations without derivatives.
This is because $\Xi$ is a twisted chiral superfield in which the signs of the derivative terms of the component fields are flipped compared with those of the 
prepotential $C$, or those of the original chiral superfield $\Phi$.
On the other hand, there are no equations for $y^2$ without derivatives.
Then we have to utilize (\ref{SUSY-d+y2-2}) and (\ref{SUSY-d-y2-2}).
Indeed, it has been recognized that they carry the information of the duality transformation \cite{Kimura:2013fda}. 
As a conclusion, we obtain the additional variation $\delta' \Xi$, and the total supersymmetry transformation of $\Xi$ is defined by the sum of (\ref{C2Xi}) and the original variation (\ref{SUSY-tchiral}) whose component fields are given by (\ref{PhiQtQXiPsi}).
The procedure of the modification $\delta' \Xi$ is completely parallel to that of a charged chiral superfield coupled to a vector superfield under the Wess-Zumino gauge.

We should also point that the absence of $F'_X$ is significant.
This is because $F'_X$ with $\theta^+ \theta^-$ comes from the chiral superfield $X'$ in the superfield formalism.
Generically, however, there are no counter terms in the twisted chiral superfield $\Xi$. 
This implies that any non-vanishing term of $F'_X$ cannot be absorbed into the modified supersymmetry transformation of $\Xi$.
Fortunately, we can find an appropriate gauge-fixing (\ref{FGN-2}), the supersymmetry transformation (\ref{SUSY-C-XYY}), and the modification (\ref{0-cond-1'}) in which the term proportional to $\theta^+ \theta^-$ is never generated. 

%%%%%%%%%%%%%%%%%%%%%%%%%%%%%%%
\subsubsection*{Additional variation $\delta' \Psi$}

We move on to the modification $\delta' \Psi$, which is originated from the additional variation $\delta' C'$. 
They are subject to the equation (\ref{SUSY'-PsiC}). 
This is expressed in terms of the component fields as
\begin{align}
(D'_c - \ol{D}{}'_c) \, \delta' r^2 
\ &= \ 
- (\delta' D'_c - \delta' \ol{D}{}'_c) \, r^2 
\nn \\
\ &= \ 
\frac{1}{2} \, \Big\{
\big( \phi'_X - \sigma'_Y + \I \ol{\sigma}{}'_Y \big)
- \big( \ol{\phi}{}'_X - \ol{\sigma}{}'_Y - \I \sigma'_Y \big)
\Big\}
\, \del_+ \del_- r^2 
\, . 
\end{align}
There are no other modification of the component field of $\Psi$.
As we have already mentioned in section \ref{sect:GF-CF}, the complex scalar $D'_c$ should not vanish in order to obtain the appropriate sigma model for the exotic five-brane.
Then we obtain 
\begin{align}
\delta' r^2 
\ &= \ 
\frac{1}{2 (D'_c - \ol{D}{}'_c)} \, \Big\{
\big( \phi'_X - \sigma'_Y + \I \ol{\sigma}{}'_Y \big)
- \big( \ol{\phi}{}'_X - \ol{\sigma}{}'_Y - \I \sigma'_Y \big)
\Big\}
\, \del_+ \del_- r^2 
\, . \label{C2Psi}
\end{align}
We note that the scalar field $r^2$ is not a dynamical field in the sigma model for the exotic five-brane \cite{Kimura:2013fda}.
The dynamical feature is carried by the scalar field $y^2$ via the duality relation (\ref{Psi2Xi}) discussed in appendix \ref{app:GLSM}.

%%%%%%%%%%%%%%%%%%%%%%%%%%%%%%%%%%%%%%%%%%%%%%%%%%%%%%%%%%%%%%%%%%%%%%%
%\newpage
\section{Summary} 
\label{sect:summary}

In this work we have studied the supergauge transformation of the prepotential $C$ 
of the chiral superfield via $\Phi = \ol{D}{}_+ \ol{D}{}_- C$.
By definition, this supergauge transformation is irrelevant for the chiral superfield.
Since the prepotential carries many redundant component fields, 
we have tried to remove some of them by the supergauge transformation.
Indeed this situation is parallel to that of a vector superfield in supersymmetric theory with four supercharges.
It is noticed that the complex scalar $D_c$ and the real part of two vectorial fields $(A_{c=}, B_{c\+})$ should not be gauged away, because they play a crucial role in generating the nongeometric structure which we have already understood in the previous works \cite{Kimura:2013fda}.
If we adopt a complex linear superfield $L$ as the supergauge parameter, the above important component fields are removed and we cannot obtain the correct sigma model.
Then we have imposed a relaxed gauge-fixing condition rather than the full gauge-fixing condition by the complex linear superfield.

The resultant gauge-fixed prepotential
$C'$ involves twice as many degrees of freedom as the original chiral superfield $\Phi$.
We have noted that $C'$ is neither a left semichiral superfield nor a right semichiral superfield which describe a sigma model whose target space is a generalized K\"{a}hler geometry.
Thus it turns out that not only the semichiral superfields but also $C'$ play a central role in generating the nongeometric structure of string theory.
We have noticed that $C'$, or the original chiral superfield $\Phi$, belongs to an $\N=(4,4)$ vector multiplet, whilst semichiral superfields are building blocks of matter supermultiplets.

The supersymmetry transformation breaks the gauge-fixing condition. 
Thus a modification of the supersymmetry transformation which does not prevent the gauge-fixing condition should be considered.
We have found a consistent modification rule as a set of linear differential equations, though it is not necessary to obtain the explicit solution. 
The prepotential $C$ couples to matter supermultiplets in the GLSM.
Thus, once the gauge-fixing condition is imposed, 
the supersymmetry transformation of the matter superfields should be modified.
We have also studied a modification rule which does not prevent the nongeometric structure of the target space geometry of the IR sigma model.

The supergauge symmetry and its gauge-fixing will be important when we carefully study quantum structure of the GLSM for the exotic five-brane.
We have studied the worldsheet instanton corrections to the exotic five-brane in the language of the gauge theory vortex corrections in \cite{Kimura:2013zva} based on the works \cite{Ooguri:1996me} and \cite{Tong:2002rq, Harvey:2005ab, Okuyama:2005gx}.
In the next step, it would be interesting to investigate the quantum structure coming from the prepotential
which we have not seriously analyzed in the previous work \cite{Kimura:2013zva}.

%%%%%%%%%%%%%%%%%%%%%%%%%%%%%%%%%%%%%%%%%%%%%%%%%%%%%%%%%%%%%%%%%%%%%%%
%\newpage
\section*{Acknowledgments}

The author thanks
Goro Ishiki,
Hisayoshi Muraki,
Akio Sugamoto,
Masato Taki
and
Akinori Tanaka
for valuable comments.
This work is supported by the MEXT-Supported Program for the Strategic Research Foundation at Private Universities ``Topological Science'' (Grant No.~S1511006). 
This is also supported in part by the Iwanami-Fujukai Foundation.

%%%%%%%%%%%%%%%%%%%%%%%%%%%%%%%%%%%%%%%%%%%%%%%%%%%%%%%%%%%%%%%%%%%%%%%
%\newpage
\begin{appendix}

\section*{Appendix}

%%%%%%%%%%%%%%%%%%%%%%%%%%%%%%%%%%%%%%%%%%%%%%%%%%%%%%%%%%%%%%%%%%%%%%%
\section{Conventions}
\label{app:conventions}

It is useful to exhibit conventions and notations of supersymmetric field theory in two-dimensional spacetime with the Lorentz signature.

%%%%%%%%%%%%%%%
\subsection{Weyl spinors}

Following \cite{Witten:1993yc}, we introduce the convention of Weyl spinor indices:
\bsubeq
\begin{gather}
(\theta^1, \theta^2) \ \equiv \ (\theta^-, \theta^+)
\, , \ls
(\theta^{\alpha})^{\dagger} \ = \ \ol{\theta}{}^{\dot{\alpha}}
\, , \ls
(\theta^{\pm})^{\dagger} \ = \ \ol{\theta}{}^{\pm}
\, , \\
%%%%%%%%%%%%
\theta_{\alpha} \ = \ \ve_{\alpha \beta} \, \theta^{\beta}
\, , \ls
\theta^{\alpha} \ = \ \ve^{\alpha \beta} \, \theta_{\beta}
\, , \\
\ve^{-+} \ = \ \ve_{+-} \ = \ + 1
\, , \ls
\theta^- \ = \ + \theta_+ 
\, , \ls
\theta^+ \ = \ - \theta_-
\, .
\end{gather}
\esubeq
In the superspace formalism,
supercovariant derivatives $D_{\alpha}$, $\ol{D}_{\dot{\alpha}}$
and supercharges $Q_{\alpha}$, $\ol{Q}_{\dot{\alpha}}$ in two-dimensional spacetime are written as 
\bsubeq
\begin{align}
D_{\pm} \ &= \ 
\frac{\del}{\del \theta^{\pm}} 
- \I \ol{\theta}{}^{\pm} \big( \del_0 \pm \del_1 \big)
\, , &
\ol{D}{}_{\pm} \ &= \ 
- \frac{\del}{\del \ol{\theta}{}^{\pm}} 
+ \I \theta^{\pm} \big( \del_0 \pm \del_1 \big)
\, , \label{covD} \\
%%%%%%%%%%%%%%%%%
Q_{\pm} \ &= \ 
\frac{\del}{\del \theta^{\pm}} 
+ \I \ol{\theta}{}^{\pm} \big( \del_0 \pm \del_1 \big)
\, , &
\ol{Q}{}_{\pm} \ &= \ 
- \frac{\del}{\del \ol{\theta}{}^{\pm}} 
- \I \theta^{\pm} \big( \del_0 \pm \del_1 \big)
\, . \label{SUSY-Q}
\end{align}
\esubeq
Here $\theta^{\pm}$ and $\ol{\theta}{}^{\pm}$ are the Grassmann coordinates.
We also define their integral measures,
\bsubeq \label{f-measure-22}
\begin{gather}
\d^2 \theta \ = \ 
- \frac{1}{4} \, \d \theta^{\alpha} \, \d \theta^{\beta} \,
\ve_{\alpha \beta} 
\ = \ 
- \half \, \d \theta^+ \, \d \theta^- \, , \ls
\d^2 \ol{\theta} \ = \ 
- \frac{1}{4} \, \d \ol{\theta}{}_{\dot{\alpha}} \, 
\d \ol{\theta}{}_{\dot{\beta}} \, \ve^{\dot{\alpha} \dot{\beta}} 
\ = \ 
\half \, \d \ol{\theta}{}^+ \, \d \ol{\theta}{}^- \, , \\
\d^2 \wt{\theta} \ = \ 
- \half \, \d \theta^+ \, \d \ol{\theta}{}^- \, , \ls 
\d^2 \ol{\wt{\theta}} \ = \ 
- \half \, \d \theta^- \, \d \ol{\theta}{}^+ \, , \\
\d^4 \theta \ = \ \d^2 \theta \, \d^2 \ol{\theta} 
\ = \ 
- \d^2 \wt{\theta} \, \d^2 \ol{\wt{\theta}} 
\ = \ - \frac{1}{4} \d \theta^+ \, \d \theta^- \, \d \ol{\theta}{}^+ \,
\d \ol{\theta}{}^- \, .
\end{gather}
We note that the definition of hermitian conjugate is $(\eta_+ \lambda_-)^{\dagger} = + \ol{\lambda}{}_- \ol{\eta}{}_+$.
Then integrals of $(\theta^{\pm}, \ol{\theta}{}^{\pm})$ are given as 
\begin{align}
\int \! \d^2 \theta \, \theta^+ \theta^- 
\ &= \ \half 
\, , \ls
\int \! \d^2 \wt{\theta} \, \theta^+ \ol{\theta}{}^- 
\ = \ 
\half 
\, . %\label{f-int}
\end{align}
\esubeq

%%%%%%%%%%%%%%%
\subsection{Superfields}
\label{app:SF}

In two-dimensional spacetime, 
there exist various irreducible supersymmetric multiplets.
In this paper we focus on a chiral multiplet $(\phi_X, \psi_{X\pm}, F_X)$
and a twisted chiral multiplet $(\sigma_Y, \ol{\chi}{}_{Y+}, \chi_{Y-}, G_Y)$.
Here $\phi_X$ and $\sigma_Y$ are complex scalar fields,
$\psi_{X\pm}$, $\ol{\chi}{}_{Y+}$ and $\chi_{Y-}$ are Weyl spinors.
They are dynamical.
We also introduce auxiliary complex scalar fields $F_X$ and $G_Y$.
These two supermultiplets can be expressed as superfields $X$ and $Y$
whose definitions are 
$0 = \ol{D}{}_{\pm} X$ and 
$0 = \ol{D}{}_+ Y = D_- Y$,
respectively.
Their expansions in terms of $(\theta^{\pm}, \ol{\theta}{}^{\pm})$ are defined as 
\bsubeq \label{expand}
\begin{align}
X \ &= \
\phi_X
+ \I \sqrt{2} \, \theta^+ \psi_{X+} 
+ \I \sqrt{2} \, \theta^- \psi_{X-}
+ 2 \I \, \theta^+ \theta^- F_X
\nn \\
\ & \ \ \ \ 
- \I \, \theta^+ \ol{\theta}{}^+ \del_+ \phi_X
- \I \, \theta^- \ol{\theta}{}^- \del_- \phi_X
+ \sqrt{2} \, \theta^+ \ol{\theta}{}^+ \theta^- \del_+ \psi_{X-}
+ \sqrt{2} \, \theta^- \ol{\theta}{}^- \theta^+ \del_- \psi_{X+}
\nn \\
\ & \ \ \ \
+ \theta^+ \theta^- \ol{\theta}{}^+ \ol{\theta}{}^- \del_+ \del_- \phi_X
\, , \label{chiralX} \\
%%%%%%%%%%%%%%%%
Y \ &= \
\sigma_Y
+ \I \sqrt{2} \, \theta^+ \ol{\chi}{}_{Y+} 
- \I \sqrt{2} \, \ol{\theta}{}^- \chi_{Y-} 
+ 2 \I \, \theta^+ \ol{\theta}{}^- G_Y
\nn \\
\ & \ \ \ \ 
- \I \, \theta^+ \ol{\theta}{}^+ \del_+ \sigma_Y
+ \I \, \theta^- \ol{\theta}{}^- \del_- \sigma_Y
- \sqrt{2} \, \theta^- \ol{\theta}{}^- \theta^+ \del_- \ol{\chi}{}_{Y+}
- \sqrt{2} \, \theta^+ \ol{\theta}{}^+ \ol{\theta}{}^- \del_+ \chi_{Y-}
\nn \\
\ & \ \ \ \ 
- \theta^+ \theta^- \ol{\theta}{}^+ \ol{\theta}{}^- \del_+ \del_- \sigma_Y
\, . \label{tchiralY}
\end{align}
\esubeq
Here, for convenience, we used the following derivatives of the spacetime coordinates $(x^0, x^1)$:
\begin{align}
\del_{\pm} 
\ &\equiv \ 
\del_0 \pm \del_1
\ = \ 
\frac{\del}{\del x^0} \pm \frac{\del}{\del x^1}
\, . \label{LC-deriv}
\end{align}

In the main part of this paper we study 
the prepotential $C$ of the chiral superfield $\Phi$.
We explicitly express the expansion\footnote{We notice that the convention is slightly different from that in \cite{Kimura:2013fda}.} of $C$ by the Grassmann coordinates $(\theta^{\pm}, \ol{\theta}{}^{\pm})$,
\begin{align}
C \ &= \ 
\phi_{c} 
+ \I \sqrt{2} \, \theta^+ \psi_{c+} 
+ \I \sqrt{2} \, \theta^- \psi_{c-} 
+ \I \sqrt{2} \, \ol{\theta}{}^+ \chi_{c+} 
+ \I \sqrt{2} \, \ol{\theta}{}^- \chi_{c-}
\nn \\
\ & \ \ \ \ 
+ \I \, \theta^+ \theta^- F_{c} 
+ \I \, \ol{\theta}{}^+ \ol{\theta}{}^- M_{c}
+ \theta^+ \ol{\theta}{}^- G_{c} 
+ \theta^- \ol{\theta}{}^+ N_{c}
%\nn \\
%
%\ & \ \ \ \ 
+ \theta^- \ol{\theta}{}^- A_{c=}
+ \theta^+ \ol{\theta}{}^+ B_{c\+}
\nn \\
\ & \ \ \ \ 
- \sqrt{2} \, \theta^+ \theta^- \ol{\theta}{}^+ \zeta_{c+}
- \sqrt{2} \, \theta^+ \theta^- \ol{\theta}{}^- \zeta_{c-}
- \sqrt{2} \, \theta^+ \ol{\theta}{}^+ \ol{\theta}{}^- \lambda_{c+}
- \sqrt{2} \, \theta^- \ol{\theta}{}^+ \ol{\theta}{}^- \lambda_{c-}
\nn \\
\ & \ \ \ \ 
- 2 \theta^+ \theta^- \ol{\theta}{}^+ \ol{\theta}{}^- D_{c}
\, . 
\end{align}
Here $(\phi_c, F_c, M_c, G_c, N_c, D_c)$ are complex scalars,
$(\psi_{c\pm}, \chi_{c\pm}, \zeta_{c\pm}, \lambda_{c\pm})$ are Weyl spinors.
We also introduce $(A_{c=}, B_{c\+})$, which behave as light-cone components of a complex vector field $W_{c,m}$ in such a way that $A_{c=} = \half (W_{c,0} - W_{c,1})$ and $B_{c\+} = \half (W_{c,0} + W_{c,1})$, 
though we do not explicitly use $W_{c,m}$ in this paper.

We also describe the expansion of a complex linear superfield $L$ 
defined by $0 = \ol{D}{}_+ \ol{D}{}_- L$ \cite{Gates:1984nk, Grisaru:1997ep}:
\begin{align}
L \ &= \ 
\phi_{L} 
+ \I \sqrt{2} \, \theta^+ \psi_{L+} 
+ \I \sqrt{2} \, \theta^- \psi_{L-} 
+ \I \sqrt{2} \, \ol{\theta}{}^+ \chi_{L+} 
+ \I \sqrt{2} \, \ol{\theta}{}^- \chi_{L-}
\nn \\
\ & \ \ \ \ 
+ \I \, \theta^+ \theta^- F_{L} 
+ \theta^+ \ol{\theta}{}^- G_{L} 
+ \theta^- \ol{\theta}{}^+ N_{L}
+ \theta^- \ol{\theta}{}^- A_{L=}
+ \theta^+ \ol{\theta}{}^+ B_{L\+}
\nn \\
\ & \ \ \ \ 
- \sqrt{2} \, \theta^+ \theta^- \ol{\theta}{}^+ \zeta_{L+}
- \sqrt{2} \, \theta^+ \theta^- \ol{\theta}{}^- \zeta_{L-}
+ \sqrt{2} \, \theta^+ \ol{\theta}{}^+ \ol{\theta}{}^- \del_+ \chi_{L-}
- \sqrt{2} \, \theta^- \ol{\theta}{}^+ \ol{\theta}{}^- \del_- \chi_{L+}
\nn \\
\ & \ \ \ \ 
+ \theta^+ \theta^- \ol{\theta}{}^+ \ol{\theta}{}^- \Big(
  \I \del_- B_{L\+} 
+ \I \del_+ A_{L=} 
- \del_+ \del_- \phi_{L} 
\Big)
\, . \label{CLS}
\end{align}
This superfield is reducible and carries six complex bosons 
$(\phi_L, F_L, G_L, N_L, A_{L=}, B_{L\+})$
and six Weyl fermions $(\psi_{L\pm}, \chi_{L\pm}, \zeta_{L\pm})$.

%%%%%%%%%%%%%%%%%%%%%%%%%%%%%%%%%%%%%%%%%%%%%%%%%%%%%%%%%%%%%%%%%%%%%%%
\section{Supersymmetry transformations}
\label{app:SUSYtr}

In this appendix we summarize the supersymmetry transformations of supersymmetric multiplets.
In general, we define the supersymmetry transformations of a general 
complex superfield ${\cal F}$ in terms of the supercharges $Q_{\pm}$ and $\ol{Q}{}_{\pm}$ defined in (\ref{SUSY-Q}):
\begin{align}
\delta {\cal F} \ &\equiv \ 
\Big( - \I \ve_- Q_+ + \I \ve_+ Q_- - \I \ol{\ve}{}_- \ol{Q}{}_+ + \I \ol{\ve}{}_+ \ol{Q}{}_- \Big) {\cal F}
\, . \label{SUSY-tr}
\end{align}
Here $\ve_{\pm}$ and $\ol{\ve}{}_{\pm}$ are supersymmetry parameters given as anti-commuting Weyl spinors.

First, we exhibit the transformations of the chiral multiplet $X$:
\bsubeq \label{SUSY-chiral}
\begin{align}
\delta \phi_X 
\ &= \ 
\sqrt{2} \, \ve_- \psi_{X+} 
- \sqrt{2} \, \ve_+ \psi_{X-}
\, , \\
%%%%%%%%%%%%%
\delta F_X 
\ &= \ 
- \sqrt{2} \, \ol{\ve}{}_- \del_+ \psi_{X-}
- \sqrt{2} \, \ol{\ve}{}_+ \del_- \psi_{X+}
\, , \\
%%%%%%%%%%%%%
\delta \psi_{X+} 
\ &= \ 
- \I \sqrt{2} \, \ol{\ve}{}_- \del_+ \phi_X 
+ \I \sqrt{2} \, \ve_+ F_X
\, , \\
%%%%%%%%%%%%%
\delta \psi_{X-} 
\ &= \ 
  \I \sqrt{2} \, \ol{\ve}{}_+ \del_- \phi_X
+ \I \sqrt{2} \, \ve_- F_X 
\, .
\end{align}
\esubeq
Second, the transformations of the twisted chiral multiplet $Y$ are listed as
\bsubeq \label{SUSY-tchiral}
\begin{align}
%%%%%%%%%%%%%
\delta \sigma_y 
\ &= \ 
  \sqrt{2} \, \ve_- \ol{\chi}{}_{Y+} 
- \sqrt{2} \, \ol{\ve}{}_+ \chi_{Y-} 
\, , \\
%%%%%%%%%%%
\delta G_Y 
\ &= \ 
  \sqrt{2} \, \ve_+ \del_- \ol{\chi}{}_+
+ \sqrt{2} \, \ol{\ve}{}_- \del_+ \chi_{Y-}
\, , \\
%%%%%%%%%%%
\delta \ol{\chi}{}_{Y+} 
\ &= \ 
- \I \sqrt{2} \, \ol{\ve}{}_- \del_+ \sigma_Y
- \I \sqrt{2} \, \ol{\ve}{}_+ G_Y
\, , \\
%%%%%%%%%%%
\delta \chi_{Y-} 
\ &= \ 
  \I \sqrt{2} \, \ve_+ \del_- \sigma_Y
- \I \sqrt{2} \, \ve_- G_Y
\, .
\end{align}
\esubeq
We also express the transformations of the multiplet $C$ {\it before} imposing a gauge-fixing condition. 
They are given by all the component fields:
\bsubeq \label{SUSY-C-BGF}
\begin{align}
\delta \phi_c
\ &= \ 
  \sqrt{2} \, \ve_- \psi_{c+} 
- \sqrt{2} \, \ve_+ \psi_{c-} 
- \sqrt{2} \, \ol{\ve}{}_- \chi_{c+} 
+ \sqrt{2} \, \ol{\ve}{}_+ \chi_{c-}
\, , \\
%%%%%%%%%%%%%%%%%%%
\delta F_{c}
\ &= \ 
- \sqrt{2} \, \ol{\ve}{}_- \big(
  \zeta_{c+}
+ \del_+ \psi_{c-} 
\big)
+ \sqrt{2} \, \ol{\ve}{}_+ \big(
  \zeta_{c-}
- \del_- \psi_{c+} 
\big)
\, , \\
%%%%%%%%%%%%%%%%%%%
\delta M_{c}
\ &= \ 
+ \sqrt{2} \, \ve_- \big(
  \lambda_{c+}
+ \del_+ \chi_{c-}
\big)
- \sqrt{2} \, \ve_+ \big(
  \lambda_{c-}
- \del_- \chi_{c+} 
\big)
\, , \\
%%%%%%%%%%%%%%%%%%%
\delta G_{c}
\ &= \ 
+ \I \sqrt{2} \, \ve_+ \big( 
  \zeta_{c-}
+ \del_- \psi_{c+} 
\big)
+ \I \sqrt{2} \, \ol{\ve}{}_- \big(
  \lambda_{c+}
- \del_+ \chi_{c-}  
\big)
\, , \\
%%%%%%%%%%%%%%%%%%%
\delta N_{c}
\ &= \ 
+ \I \sqrt{2} \, \ve_- \big( 
  \zeta_{c+}
- \del_+ \psi_{c-} 
\big)
+ \I \sqrt{2} \, \ol{\ve}{}_+ \big(
  \lambda_{c-}
+ \del_- \chi_{c+} 
\big)
\, , \\
%%%%%%%%%%%%%%%%%%%
\delta A_{c=}
\ &= \ 
+ \I \sqrt{2} \, \ve_+ \del_- \psi_{c-} 
+ \I \sqrt{2} \, \ol{\ve}{}_+ \del_- \chi_{c-}
+ \I \sqrt{2} \, \ve_- \zeta_{c-}
+ \I \sqrt{2} \, \ol{\ve}{}_- \lambda_{c-}
\, , \\
%%%%%%%%%%%%%%%%%%%
\delta B_{c\+}
\ &= \ 
- \I \sqrt{2} \, \ve_- \del_+ \psi_{c+} 
- \I \sqrt{2} \, \ol{\ve}{}_- \del_+ \chi_{c+} 
+ \I \sqrt{2} \, \ve_+ \zeta_{c+}
+ \I \sqrt{2} \, \ol{\ve}{}_+ \lambda_{c+}
\, , \\
%%%%%%%%%%%%%%%%%%%
\delta D_{c}
\ &= \ 
  \frac{1}{\sqrt{2}} \, \ve_- \del_+ \zeta_{c-}
+ \frac{1}{\sqrt{2}} \, \ve_+ \del_- \zeta_{c+}
- \frac{1}{\sqrt{2}} \, \ol{\ve}{}_- \del_+ \lambda_{c-}
- \frac{1}{\sqrt{2}} \, \ol{\ve}{}_+ \del_- \lambda_{c+}
\, , \\
%%%%%%%%%%%%%%%%%%%
\delta \psi_{c+}
\ &= \ 
+ \frac{1}{\sqrt{2}} \, \ol{\ve}{}_- \big( 
  B_{c\+}
- \I \del_+ \phi_{c} 
\big)
+ \frac{\I}{\sqrt{2}} \, \ve_+ F_{c} 
- \frac{1}{\sqrt{2}} \, \ol{\ve}{}_+ G_{c} 
\, , \\
%%%%%%%%%%%%%%%%%%%
\delta \psi_{c-}
\ &= \
- \frac{1}{\sqrt{2}} \, \ol{\ve}{}_+ \big(
  A_{c=}
- \I \del_- \phi_{c} 
\big)
+ \frac{\I}{\sqrt{2}} \, \ve_- F_{c} 
+ \frac{1}{\sqrt{2}} \, \ol{\ve}{}_- N_{c}
\, , \\
%%%%%%%%%%%%%%%%%%%
\delta \chi_{c+}
\ &= \
+ \frac{1}{\sqrt{2}} \, \ve_- \big(
  B_{c\+}
+ \I \del_+ \phi_{c} 
\big)
- \frac{1}{\sqrt{2}} \, \ve_+ N_{c}
- \frac{\I}{\sqrt{2}} \, \ol{\ve}{}_+ M_{c}
\, , \\
%%%%%%%%%%%%%%%%%%%
\delta \chi_{c-}
\ &= \
- \frac{1}{\sqrt{2}} \, \ve_+ \big( 
  A_{c=}
+ \I \del_- \phi_{c} 
\big)
+ \frac{1}{\sqrt{2}} \, \ve_- G_{c} 
- \frac{\I}{\sqrt{2}} \, \ol{\ve}{}_- M_{c}
\, , \\
%%%%%%%%%%%%%%%%%%%
\delta \zeta_{c+}
\ &= \ 
- \frac{1}{\sqrt{2}} \, \ol{\ve}{}_+ \big(
  \del_- B_{c\+}
+ 2 \I D_{c}
\big)
+ \frac{\I}{\sqrt{2}} \, \ve_- \del_+ F_{c} 
- \frac{1}{\sqrt{2}} \, \ol{\ve}{}_- \del_+ N_{c}
\, , \\
%%%%%%%%%%%%%%%%%%%
\delta \zeta_{c-}
\ &= \ 
- \frac{1}{\sqrt{2}} \, \ol{\ve}{}_- \big( 
  \del_+ A_{c=}
+ 2 \I D_{c}
\big)
- \frac{\I}{\sqrt{2}} \, \ve_+ \del_- F_{c} 
- \frac{1}{\sqrt{2}} \, \ol{\ve}{}_+ \del_- G_{c} 
\, , \\
%%%%%%%%%%%%%%%%%%%
\delta \lambda_{c+}
\ &= \ 
- \frac{1}{\sqrt{2}} \, \ve_+ \big( 
  \del_- B_{c\+}
- 2 \I D_{c}
\big)
- \frac{1}{\sqrt{2}} \, \ve_- \del_+ G_{c} 
- \frac{\I}{\sqrt{2}} \, \ol{\ve}{}_- \del_+ M_{c}
\, , \\
%%%%%%%%%%%%%%%%%%%
\delta \lambda_{c-}
\ &= \ 
- \frac{1}{\sqrt{2}} \, \ve_- \big(
  \del_+ A_{c=}
- 2 \I D_{c}
\big)
- \frac{1}{\sqrt{2}} \, \ve_+ \del_- N_{c}
+ \frac{\I}{\sqrt{2}} \, \ol{\ve}{}_+ \del_- M_{c}
\, .
\end{align}
\esubeq

%%%%%%%%%%%%%%%%%%%%%%%%%%%%%%%%%%%%%%%%%%%%%%%%%%%%%%%%%%%%%%%%%%%%%%%
\section{Lagrangian with prepotential}
\label{app:GLSM}

In this appendix we write down three Lagrangians which include the 
prepotential $C$, or the original adjoint chiral superfield $\Phi$ in the GLSM for an exotic five-brane \cite{Kimura:2013fda}.
For simplicity, we only consider a single $U(1)$ gauge symmetry in the GLSM.

%%%%%%%%%%%%%%%%%%%%%%%%%%%%%%%%%%
\subsection{Superfields}

The first Lagrangian is given by the original adjoint chiral superfield $\Phi$ in such a way that
\begin{align}
\Scr{L}_{\Phi}
\ &= \ 
\int \d^4 \theta \, \frac{1}{e^2} |\Phi|^2
+ \Big\{
\sqrt{2} \int \d^2 \theta \, \Big( - \wt{Q} \Phi Q + s \Phi \Big)
+ \text{(h.c.)}
\Big\}
\, . \label{L-Phi}
\end{align}
Here, $e$ and $s$ are the gauge coupling constant and the complex FI parameter, respectively.
The adjoint chiral superfield $\Phi$ is coupled to charged chiral superfields $Q$ and $\wt{Q}$.
This should be rewritten in terms of the component fields of $C$.
The second is the Lagrangian of a twisted chiral superfield $\Xi$:
\begin{align}
\Scr{L}_{\Xi}
\ &= \ 
- \frac{g^2}{2} \int \d^4 \theta \, \Big(
\Xi + \ol{\Xi} 
- \sqrt{2} \, (C + \ol{C})
\Big)^2
\, . \label{L-Xi}
\end{align}
Here $g$ is the sigma model coupling constant.
The third is the Lagrangian which contains the imaginary part of $C$ coupled to the imaginary part of a chiral superfield $\Psi$:
\begin{align}
\Scr{L}_{\Psi C}
\ &= \ 
- \sqrt{2} \int \d^4 \theta \, (\Psi - \ol{\Psi}) (C - \ol{C}) 
\, . \label{L-PsiC}
\end{align}
This plays a central role in describing the exotic (i.e., nongeometric) structure of the five-brane.
Indeed the chiral superfield $\Psi$ is dual to the twisted chiral superfield $\Xi$.
We note that $\Xi$ is dynamical, while $\Psi$ is non-dynamical in this system \cite{Kimura:2013fda}.

We discuss the Lagrangians 
(\ref{L-Phi}),
(\ref{L-Xi}) and
(\ref{L-PsiC}) in terms of component fields.
In order for that, we expand the superfields $\Phi$, $Q$, $\wt{Q}$, $\Xi$ and $\Psi$ in the above Lagrangians:
\bsubeq \label{PhiQtQXiPsi}
\begin{align}
\Phi \ &= \ 
\phi
+ \I \sqrt{2} \, \theta^+ \wt{\lambda}_{+} 
+ \I \sqrt{2} \, \theta^- \wt{\lambda}_{-}
+ 2 \I \, \theta^+ \theta^- D_{\Phi}
+ \ldots 
\, , \\
%%%%%%%%%%%%%%%
Q \ &= \ 
q 
+ \I \sqrt{2} \, \theta^+ \psi_+ 
+ \I \sqrt{2} \, \theta^- \psi_-
+ 2 \I \, \theta^+ \theta^- F
+ \ldots 
\, , \\
%%%%%%%%%%%%%%%
\wt{Q} \ &= \ 
\wt{q}
+ \I \sqrt{2} \, \theta^+ \wt{\psi}_+
+ \I \sqrt{2} \, \theta^- \wt{\psi}_-
+ 2 \I \, \theta^+ \theta^- \wt{F}
+ \ldots 
\, , \\
%%%%%%%%%%%%%%%
\Xi \ &\equiv \ 
\frac{1}{\sqrt{2}} (y^1 + \I y^2)
+ \I \sqrt{2} \, \theta^+ \ol{\xi}{}_+
- \I \sqrt{2} \, \ol{\theta}{}^- \xi_-
+ 2 \I \, \theta^+ \ol{\theta}{}^- G_{\Xi}
+ \ldots 
\, , \\
%%%%%%%%%%%%%%%
\Psi \ &= \ 
\frac{1}{\sqrt{2}} (r^1 + \I r^2)
+ \I \sqrt{2} \, \theta^+ \chi_+
+ \I \sqrt{2} \, \theta^- \chi_-
+ 2 \I \, \theta^+ \theta^- G
+ \ldots 
\, .
\end{align}
\esubeq
where the symbol ``$\ldots$'' implies derivative terms governed by the supercovariant derivatives $D_{\pm}$ and $\ol{D}{}_{\pm}$ defined in (\ref{covD}).
The explicit forms can be seen in the same way as in (\ref{expand}).
Since the adjoint chiral superfield $\Phi$ is given by the prepotential $C$, 
we see the relations among their component fields in (\ref{Phi2C-2}).
It is also important to mention the duality relation between $\Xi$ and $\Psi$ (for detailed discussions, see \cite{Kimura:2013fda}):
\bsubeq \label{Psi2Xi}
\begin{align}
\Psi + \ol{\Psi}
\ &= \ 
- g^2 (\Xi + \ol{\Xi})
+ \sqrt{2} \, g^2 (C + \ol{C})
\, ,
\end{align}
and the duality relations among their component fields,
\begin{align}
r^1 \ &= \ 
- g^2 y^1 + g^2 (\phi_c + \ol{\phi}{}_c)
\, , \\
%%%%%%%%%%%%%%
\chi_{\pm} \ &= \ 
\mp g^2 \ol{\xi}{}_{\pm}
+ \sqrt{2} \, g^2 (\psi_{c\pm} + \ol{\chi}{}_{c\pm})
\, , \\
%%%%%%%%%%%%%%
%\ol{\chi}{}_{\pm} \ &= \ 
%\mp g^2 \xi_{\pm}
%+ \sqrt{2} \, g^2 (\ol{\psi}{}_{c\pm} + \chi_{c\pm})
%\, , \\
%%%%%%%%%%%%%%
G \ &= \ 
\frac{g^2}{\sqrt{2}} \, (F_c + \ol{M}{}_c)
\, , \\
%%%%%%%%%%%%%%
0 \ &= \ 
- G_{\Xi} - \frac{\I}{\sqrt{2}} \, (G_c + \ol{N}{}_c)
\, , \\
%%%%%%%%%%%%%%
\del_+ r^2 \ &= \ 
- g^2 \del_+ y^2 + g^2 (B_{c\+} + \ol{B}{}_{c\+})
\, , \label{r2+y2} \\
%%%%%%%%%%%%%%
\del_- r^2 \ &= \ 
+ g^2 \del_- y^2 + g^2 (A_{c=} + \ol{A}{}_{c=})
\, . \label{r2-y2}
\end{align}
\esubeq
We do not use the relation between $\del_{\pm} r^1$ and $\del_{\pm} y^1$ because the signs of the derivatives acting on the component fields of $\Xi$ are flipped compared with those of $\Psi$.
This phenomenon also appears in the relation between $\chi_{\pm}$ and $\ol{\xi}{}_{\pm}$.
They originate from the definition of a twisted chiral superfield.
On the other hand, there are no relations between $r^2$ and $y^2$ without derivatives.
Then we have to use the relations (\ref{r2+y2}) and (\ref{r2-y2}) with derivatives and flipped signs.
Indeed, they are nothing but the duality relations between the original field $r^2$ and the dual field $y^2$.

%%%%%%%%%%%%%%%%%%%%%%%%%%%%%%%%%%
\subsection{Component fields}

We describe the three Lagrangians in terms of the component fields.
The Lagrangian $\Scr{L}_{\Phi}$ (\ref{L-Phi}) is
\begin{align}
\Scr{L}_{\Phi}
\ &= \ 
\frac{1}{e^2} \Big\{
- |\del_m \phi|^2
+ |D_{\Phi}|^2
\Big\}
+ \frac{\I}{e^2} \Big\{
  \ol{\wt{\lambda}}{}_{+} \del_- \wt{\lambda}_{+}
+ \ol{\wt{\lambda}}{}_{-} \del_+ \wt{\lambda}_{-}
\Big\}
\nn \\
%%%
\ & \ \ \ \ 
- \I \sqrt{2} \Big\{
\phi \big( q \wt{F} + \wt{q} F \big)
+ (q \wt{q} - s) \, D_{\Phi} 
\Big\}
+ \I \sqrt{2} \Big\{
\ol{\phi} \big( \ol{q} \ol{\wt{F}} + \ol{\wt{q}} \ol{F} \big)
+ (\ol{q} \ol{\wt{q}} - \ol{s}) \, \ol{D}{}_{\Phi}
\Big\}
\nn \\
%%%
\ & \ \ \ \
- \sqrt{2} \Big\{
\phi \big( 
  \psi_{+} \wt{\psi}_{-} 
+ \wt{\psi}_{+} \psi_{-} 
\big)
+ \ol{\phi} \big( 
  \ol{\wt{\psi}}{}_{-} \ol{\psi}{}_{+} 
+ \ol{\psi}{}_{-} \ol{\wt{\psi}}{}_{+} 
\big)
\Big\}
\nn \\
%%%
\ & \ \ \ \
- \sqrt{2} \Big\{
q \big( 
  \wt{\lambda}_{+} \wt{\psi}_{-} 
+ \wt{\psi}_{+} \wt{\lambda}_{-} 
\big)
+ \ol{q} \big( 
  \ol{\wt{\psi}}{}_{-} \ol{\wt{\lambda}}{}_{+} 
+ \ol{\wt{\lambda}}{}_{-} \ol{\wt{\psi}}{}_{+} 
\big)
\Big\}
\nn \\
%%%
\ & \ \ \ \
- \sqrt{2} \Big\{
\wt{q} \big( 
  \wt{\lambda}_{+} \psi_{-} 
+ \psi_{+} \wt{\lambda}_{-} 
\big)
+ \ol{\wt{q}} \big( 
  \ol{\psi}{}_{-} \ol{\wt{\lambda}}{}_{+} 
+ \ol{\wt{\lambda}}{}_{-} \ol{\psi}{}_{+} 
\big)
\Big\}
\nn \\
%%%%%%%%%%%%%%%%%%%%%%%%%%
\ &= \ 
- \frac{1}{e^2} |\del_m M_c|^2
+ \frac{\I}{e^2} \Big\{ 
  \big( \ol{\lambda}{}_{c+} + \del_+ \ol{\chi}{}_{c-} \big) \del_- \big( \lambda_{c+} + \del_+ \chi_{c-} \big) 
+ \big( \ol{\lambda}{}_{c-} - \del_- \ol{\chi}{}_{c+} \big) \del_+ \big( \lambda_{c-} - \del_- \chi_{c+} \big) 
\Big\}
\nn \\
%%%
\ & \ \ \ \
- \sqrt{2} \, \Big\{ 
  M_c \big( q \wt{F} + \wt{q} F \big)
+ \ol{M}{}_c \big( \ol{q} \ol{\wt{F}} + \ol{\wt{q}} \ol{F} \big)
\Big\}
+ \sqrt{2} \, \Big( s \, D_c + \ol{s} \, \ol{D}{}_c \Big)
- 2 e^2 \, \big| q \wt{q} \big|^2
\nn \\
%%%
\ & \ \ \ \ 
+ \frac{1}{e^2} \Big| D_c - \sqrt{2} \, e^2 \, \ol{q} \ol{\wt{q}} \Big|^2
+ \frac{1}{4 e^2} \Big| \del_+ A_{c=} + \del_- B_{c\+} + \I \del_+ \del_- \phi_c \Big|^2
\nn \\
%%%%%%%%
\ & \ \ \ \ 
- \frac{\I}{2 e^2} \Big( D_c - \sqrt{2} \, e^2 \, \ol{q} \ol{\wt{q}} \Big)
\Big\{
\del_+ \ol{A}{}_{c=} + \del_- \ol{B}{}_{c\+} - \I \del_+ \del_- \ol{\phi}{}_c
\Big\}
\nn \\
%%%%%%%%
\ & \ \ \ \ 
+ \frac{\I}{2 e^2} \Big( \ol{D}{}_c - \sqrt{2} \, e^2 \, q \wt{q} \Big)
\Big\{
\del_+ A_{c=} + \del_- B_{c\+} + \I \del_+ \del_- \phi_c
\Big\}
\nn \\
%%%
\ & \ \ \ \
+ \I \sqrt{2} \, \Big\{
  M_c \big( 
  \psi_{+} \wt{\psi}_{-} 
+ \wt{\psi}_{+} \psi_{-} 
\big)
- \ol{M}{}_c \big( 
  \ol{\wt{\psi}}{}_{-} \ol{\psi}{}_{+} 
+ \ol{\psi}{}_{-} \ol{\wt{\psi}}{}_{+} 
\big)
\Big\}
\nn \\
%%%
\ & \ \ \ \
+ \I \sqrt{2} \, q \Big\{
  \wt{\psi}_{+} \big( \lambda_{c-} - \del_- \chi_{c+} \big) 
- \wt{\psi}_{-} \big( \lambda_{c+} + \del_+ \chi_{c-} \big) 
\Big\}
\nn \\
%%%
\ & \ \ \ \
+ \I \sqrt{2} \, \ol{q} \Big\{ 
  \ol{\wt{\psi}}{}_{+} \big( \ol{\lambda}{}_{c-} - \del_- \ol{\chi}{}_{c+} \big)
- \ol{\wt{\psi}}{}_{-} \big( \ol{\lambda}{}_{c+} + \del_+ \ol{\chi}{}_{c-} \big)
\Big\}
\nn \\
%%%
\ & \ \ \ \
+ \I \sqrt{2} \, \wt{q} \Big\{
  \psi_{+} \big( \lambda_{c-} - \del_- \chi_{c+} \big) 
- \psi_{-} \big( \lambda_{c+} + \del_+ \chi_{c-} \big)
\Big\}
\nn \\
%%%
\ & \ \ \ \
+ \I \sqrt{2} \, \ol{\wt{q}} \Big\{
  \ol{\psi}{}_{+} \big( \ol{\lambda}{}_{c-} - \del_- \ol{\chi}{}_{c+} \big) 
- \ol{\psi}{}_{-} \big( \ol{\lambda}{}_{c+} + \del_+ \ol{\chi}{}_{c-} \big)
\Big\}
\, . \label{L-PhiC-2}
\end{align}
Next we express the Lagrangian $\Scr{L}_{\Xi}$ (\ref{L-Xi}) as follows:
\begin{align}
\Scr{L}_{\Xi}
\ &= \ 
- \frac{g^2}{2} \Big\{ (\del_m y^1)^2 + (\del_m y^2)^2 \Big\}
+ \I g^2 \Big\{ \ol{\xi}{}_+ \del_- \xi_+ + \ol{\xi}{}_- \del_+ \xi_- \Big\}
\nn \\
\ & \ \ \ \ 
+ g^2 (D_{c} + \ol{D}{}_c) \Big\{ y^1 - (\phi_{c} + \ol{\phi}{}_c) \Big\}
+ \frac{g^2}{2} (\phi_{c} + \ol{\phi}{}_c) \, \del_+ \del_- y^1
- \frac{g^2}{2} (F_{c} + \ol{M}{}_c) (M_{c} + \ol{F}{}_c)
\nn \\
\ & \ \ \ \ 
+ g^2 |G_{\Xi}|^2
- \frac{\I g^2}{\sqrt{2}} \, \Big\{ 
  G_{\Xi} (N_{c} + \ol{G}{}_c)
- \ol{G}{}_{\Xi} (G_{c} + \ol{N}{}_c)
\Big\}
+ \frac{g^2}{2} (G_{c} + \ol{N}{}_c) (N_{c} + \ol{G}{}_c)
\nn \\
\ & \ \ \ \ 
+ \frac{g^2}{2} \, (A_{c=} + \ol{A}{}_{c=}) \del_+ y^2 
- \frac{g^2}{2} \, (B_{c\+} + \ol{B}{}_{c\+}) \del_- y^2 
- \frac{g^2}{2} (A_{c=} + \ol{A}{}_{c=}) (B_{c\+} + \ol{B}{}_{c\+})
\nn \\
\ & \ \ \ \ 
- \frac{\I g^2}{\sqrt{2}} \, \Big\{
  \ol{\xi}{}_+ (\lambda_{c-} + \ol{\zeta}{}_{c-})
+ \ol{\xi}{}_- (\lambda_{c+} + \ol{\zeta}{}_{c+})
\Big\}
- \frac{\I g^2}{\sqrt{2}} \, \Big\{ 
  \xi_+ (\zeta_{c-} + \ol{\lambda}{}_{c-})
+ \xi_- (\zeta_{c+} + \ol{\lambda}{}_{c+})
\Big\}
\nn \\
\ & \ \ \ \
+ \I g^2 \Big\{
  (\psi_{c+} + \ol{\chi}{}_{c+}) (\lambda_{c-} + \ol{\zeta}{}_{c-})
- (\psi_{c-} + \ol{\chi}{}_{c-}) (\lambda_{c+} + \ol{\zeta}{}_{c+})
\Big\}
\nn \\
\ & \ \ \ \ 
+ \I g^2 \Big\{
  (\chi_{c+} + \ol{\psi}{}_{c+}) (\zeta_{c-} + \ol{\lambda}{}_{c-})
- (\chi_{c-} + \ol{\psi}{}_{c-}) (\zeta_{c+} + \ol{\lambda}{}_{c+})
\Big\}
\nn \\
\ & \ \ \ \ 
- \frac{\I g^2}{\sqrt{2}} \, \Big\{ 
  \del_+ \ol{\xi}{}_- (\chi_{c-} + \ol{\psi}{}_{c-})
- \del_- \ol{\xi}{}_+ (\chi_{c+} + \ol{\psi}{}_{c+})
\Big\}
\nn \\
\ & \ \ \ \ 
- \frac{\I g^2}{\sqrt{2}} \, \Big\{
  \del_+ \xi_- (\psi_{c-} + \ol{\chi}{}_{c-})
- \del_- \xi_+ (\psi_{c+} + \ol{\chi}{}_{c+})
\Big\}
\nn \\
%%%%%%%%%%%%%%%%%%%
\ &= \ 
- \frac{1}{2 g^2} (\del_m r^1)^2
- \frac{g^2}{2} (\del_m y^2)^2 
+ \frac{\I}{g^2} \Big\{ \chi_+ \del_- \ol{\chi}{}_+ + \ol{\chi}{}_- \del_+ \chi_- \Big\}
\nn \\
\ & \ \ \ \ 
- (D_{c} + \ol{D}{}_c) r^1 
+ \half (\phi_c + \ol{\phi}{}_c) \, \del_+ \del_- r^1
- \frac{g^2}{2} (F_{c} + \ol{M}{}_c) (M_{c} + \ol{F}{}_c)
\nn \\
\ & \ \ \ \ 
+ \frac{g^2}{2} \, (A_{c=} + \ol{A}{}_{c=}) \del_+ y^2 
- \frac{g^2}{2} \, (B_{c\+} + \ol{B}{}_{c\+}) \del_- y^2 
- \frac{g^2}{2} (A_{c=} + \ol{A}{}_{c=}) (B_{c\+} + \ol{B}{}_{c\+})
\nn \\
\ & \ \ \ \ 
+ \frac{\I}{\sqrt{2}} \, \chi_+ \Big\{ 
  \big( \lambda_{c-} - \del_- \chi_{c+} \big)
+ \big( \ol{\zeta}{}_{c-} - \del_- \ol{\psi}{}_{c+} \big)
\Big\}
\nn \\
\ & \ \ \ \ 
- \frac{\I}{\sqrt{2}} \, \chi_- \Big\{
  \big( \lambda_{c+} + \del_+ \chi_{c-} \big)
+ \big( \ol{\zeta}{}_{c+} + \del_+ \ol{\psi}{}_{c-} \big)
\Big\}
\nn \\
\ & \ \ \ \ 
+ \frac{\I}{\sqrt{2}} \, \ol{\chi}{}_+ \Big\{
  \big( \ol{\lambda}{}_{c-} - \del_- \ol{\chi}{}_{c+} \big)
+ \big( \zeta_{c-} - \del_- \psi_{c+} \big)
\Big\}
\nn \\
\ & \ \ \ \ 
- \frac{\I}{\sqrt{2}} \, \ol{\chi}{}_- \Big\{
  \big( \ol{\lambda}{}_{c+} + \del_+ \ol{\chi}{}_{c-} \big)
+ \big( \zeta_{c+} + \del_+ \psi_{c-} \big)
\Big\}
\, . \label{L-Xi-2}
\end{align}
Finally the Lagrangian $\Scr{L}_{\Psi C}$ (\ref{L-PsiC}) is expressed as
\begin{align}
\Scr{L}_{\Psi C}
\ &= \ 
\frac{1}{\sqrt{2}} \Big\{
  G (\ol{F}{}_c - M_c)
+ \ol{G} (F_{c} - \ol{M}{}_c)
\Big\}
\nn \\
\ & \ \ \ \ 
+ \Big\{
  \frac{\I}{2} (\phi_{c} - \ol{\phi}{}_c) \del_+ \del_- r^2
- \I \, (D_{c} - \ol{D}{}_c) r^2
\Big\}
+ \frac{\I}{2} \Big\{
  (A_{c=} - \ol{A}{}_{c=}) \del_+ r^1 
+ (B_{c\+} - \ol{B}{}_{c\+}) \del_- r^1 
\Big\}
\nn \\
\ & \ \ \ \ 
+ \frac{\I}{\sqrt{2}} \, \chi_+ \Big\{
  \big( \lambda_{c-} - \del_- \chi_{c+} \big)
- \big( \ol{\zeta}{}_{c-} - \del_- \ol{\psi}{}_{c+} \big)
\Big\}
\nn \\
\ & \ \ \ \ 
- \frac{\I}{\sqrt{2}} \, \chi_- \Big\{
  \big( \lambda_{c+} + \del_+ \chi_{c-} \big)
- \big( \ol{\zeta}{}_{c+} + \del_+ \ol{\psi}{}_{c-} \big)
\Big\}
\nn \\
\ & \ \ \ \ 
+ \frac{\I}{\sqrt{2}} \, \ol{\chi}{}_+ \Big\{
  \big( \ol{\lambda}{}_{c-} - \del_- \ol{\chi}{}_{c+} \big)
- \big( \zeta_{c-} - \del_- \psi_{c+} \big)
\Big\}
\nn \\
\ & \ \ \ \ 
- \frac{\I}{\sqrt{2}} \, \ol{\chi}{}_- \Big\{
  \big( \ol{\lambda}{}_{c+} + \del_+ \ol{\chi}{}_{c-} \big)
- \big( \zeta_{c+} + \del_+ \psi_{c-} \big)
\Big\}
\, . \label{L-PsiC-2}
\end{align}
We note that the first equations in the right-hand sides of (\ref{L-PhiC-2}) and (\ref{L-Xi-2}) are given by the component fields of the adjoint chiral superfield $\Phi$, while their second equations are expressed by the component fields of the prepotential $C$ via the correspondence (\ref{Phi2C-2}).

%%%%%%%%%%%%%%%%%%%%%%%%%%%%%%%%%%%%%%%%%%%%%%%%%%%%%%
%\newpage
\subsection{Field equations for component fields of $C$}

Focusing on the Lagrangians (\ref{L-PhiC-2}), (\ref{L-Xi-2}) and (\ref{L-PsiC-2}),
we write down the field equations for the component fields $(D_c, F_c, A_{c=}, B_{c\+}, \phi_c)$ of the prepotential $C$. 
Here we set the complex FI parameter $s = \frac{1}{\sqrt{2}} (s^1 + \I s^2)$, while we do not impose any gauge-fixing conditions on $C$:
\bsubeq \label{EOM-aux}
\begin{alignat}{2}
D_c :& &\ \ 
0 \ &= \ 
\frac{1}{e^2} \big( \ol{D}{}_c - \sqrt{2} \, e^2 q \wt{q} \big)
- \frac{\I}{2 e^2} \Big\{ \del_+ \ol{A}{}_{c=} + \del_- \ol{B}{}_{c\+} - \I \del_+ \del_- \ol{\phi}{}_c \Big\} 
\nn \\
&&& \ \ \ \ 
- (r^1 - s^1) - \I (r^2 - s^2) 
\, , \label{EOM-Dc} \\
%%%%%%%%%%%%%%%%%%
F_c :& &\ \ 
0 \ &= \ 
- \frac{g^2}{2} (M_c + \ol{F}{}_c)
- \frac{1}{\sqrt{2}} \ol{G}
\, , \\
%%%%%%%%%%%%%%%%%%
A_{c=} :& &\ \ 
0 \ &= \ 
- \frac{\I}{2 e^2} \del_+ \big( \ol{D}{}_c - \sqrt{2} \, e^2 q \wt{q} \big)
- \frac{1}{4 e^2} \del_+ \Big\{ \del_+ \ol{A}{}_{c=} + \del_- \ol{B}{}_{c\+} - \I \del_+ \del_- \ol{\phi}{}_c \Big\} 
\nn \\
&&& \ \ \ \ 
+ \frac{\I}{2} \del_+ r^1
+ \frac{g^2}{2} \Big\{ \del_+ y^2 - (B_{c\+} + \ol{B}{}_{c\+}) \Big\}
\, , \label{EOM-A} \\
%%%%%%%%%%%%%%%%%%
B_{c\+} :& &\ \ 
0 \ &= \ 
- \frac{\I}{2 e^2} \del_- \big( \ol{D}{}_c - \sqrt{2} \, e^2 q \wt{q} \big)
- \frac{1}{4 e^2} \del_- \Big\{ \del_+ \ol{A}{}_{c=} + \del_- \ol{B}{}_{c\+} - \I \del_+ \del_- \ol{\phi}{}_c \Big\} 
\nn \\
&&& \ \ \ \ 
+ \frac{\I}{2} \del_- r^1
- \frac{g^2}{2} \Big\{ \del_- y^2 + (A_{c=} + \ol{A}{}_{c=}) \Big\}
\, , \label{EOM-B} \\
%%%%%%%%%%%%%%%%%%
\phi_c :& &\ \ 
0 \ &= \
- \frac{1}{2 e^2} \del_+ \del_- \big( \ol{D}{}_c - \sqrt{2} \, e^2 q \wt{q} \big)
+ \frac{\I}{4 e^2} \del_+ \del_- \Big\{ \del_+ \ol{A}{}_{c=} + \del_- \ol{B}{}_{c\+} - \I \del_+ \del_- \ol{\phi}{}_c \Big\} 
\nn \\
&&& \ \ \ \ 
+ \half \del_+ \del_- (r^1 + \I r^2)
\, . \label{EOM-phi}
\end{alignat}
\esubeq

%%%%%%%%%%%%%%%%%%%%%%%%%%%%%%%%%%%%%%%%%%%%%%%%%%%%%%
%\newpage

\subsection{GLSM for exotic $5^2_2$-brane}

In order to recognize the importance of the auxiliary fields in the prepotential $C$, we briefly demonstrate the analysis of the GLSM for the exotic $5^2_2$-brane \cite{Kimura:2013fda}.
The Lagrangian is given by
\begin{align}
\Scr{L}
\ &= \ 
\sum_{a=1}^k \int \d^4 \theta \, \Big\{
\frac{1}{e_a^2} \Big( - |\Sigma_a|^2 + |\Phi_a|^2 \Big)
+ |Q_a|^2 \, \e^{+ 2 V_a} 
+ |\wt{Q}_a|^2 \, \e^{- 2 V_a}
\Big\}
\nn \\
\ & \ \ \ \ 
+ \int \d^4 \theta \, \frac{g^2}{2} \Big\{
- \Big( \Xi + \ol{\Xi} - \sqrt{2} \sum_{a=1}^k (C_a + \ol{C}{}_a) \Big)^2
+ \Big( \Gamma + \ol{\Gamma} + 2 \sum_{a=1}^k V_a \Big)^2
\Big\}
\nn \\
\ & \ \ \ \ 
+ \sum_{a=1}^k \Big\{ 
\sqrt{2} \int \d^2 \theta \, \big( - \wt{Q}_a \Phi_a Q_a + s_a \, \Phi_a \big)
+ \text{(h.c.)}
\Big\}
+ \sum_{a=1}^k \Big\{ 
\sqrt{2} \int \d^2 \wt{\theta} \, t_a \, \Sigma_a
+ \text{(h.c.)}
\Big\}
\nn \\
\ & \ \ \ \ 
- \sqrt{2} \int \d^4 \theta \, (\Psi - \ol{\Psi}) \sum_{a=1}^k (C_a - \ol{C}{}_a)
+ \sqrt{2} \, \ve^{mn} \sum_{a=1}^k \del_m (r^4 A_{n,a})
\, . \label{GLSM-522}
\end{align}
Here we attached the label $a$ which represents the multiplets interacting the $a$-th gauge multiplet with the $U(1)_a$ gauge group.
We focus only on the bosonic terms:
\begin{align}
\Scr{L}
\ &= \ 
\sum_a \frac{1}{e_a^2} \Big\{
\half (F_{01,a})^2
- |\del_m \sigma_a|^2
- |\del_m M_{c,a}|^2
\Big\}
- \sum_a \Big\{ 
|D_m q_a|^2
+ |D_m \wt{q}_a|^2
\Big\}
\nn \\
\ & \ \ \ \ 
- \frac{1}{2 g^2} \Big\{ (\del_m r^1)^2 + (\del_m r^3)^2 \Big\}
- \frac{g^2}{2} \Big\{ (\del_m y^2)^2 + (D_m \gamma^4)^2 \Big\}
+ \sqrt{2} \, \ve^{mn} \sum_a \del_m \big( (r^4 - t^4_a) A_{n,a} \big)
\nn \\
\ & \ \ \ \ 
- 2 g^2 \sum_{a,b} \sigma_a \ol{\sigma}{}_b 
- 2 \sum_a |\sigma_a|^2 \big( |q_a|^2 + |\wt{q}_a|^2 \big)
\nn \\
\ & \ \ \ \ 
+ \sum_a \Big\{ \frac{1}{2 e_a^2} (D_{V,a})^2
- D_{V,a} \big( |q_a|^2 - |\wt{q}_a|^2 - \sqrt{2} \, (r^3 - t^3_a) \big)
\Big\}
\nn \\
\ & \ \ \ \ 
+ \sum_a \Big\{ 
|F_a|^2 + |\wt{F}_a|^2
- \sqrt{2} M_{c,a} \big( q_a \wt{F}_a + \wt{q}_a F_a \big)
- \sqrt{2} \ol{M}{}_{c,a} \big( \ol{q}{}_a \ol{\wt{F}}{}_a + \ol{\wt{q}}{}_a \ol{F}{}_a \big)
\Big\}
+ g^2 |G_{\Gamma}|^2
\nn \\
\ & \ \ \ \ 
+ \frac{1}{\sqrt{2}} \sum_a \Big\{ (F_{c,a} - \ol{M}{}_{c,a}) \ol{G} + (\ol{F}{}_{c,a} - M_{c,a}) G \Big\}
- \frac{g^2}{2} \sum_{a,b} (F_{c,a} + \ol{M}{}_{c,a}) (\ol{F}{}_{c,b} + M_{c,b})
\nn \\
\ & \ \ \ \ 
+ {g^2} |{G}_{\Xi}|^2
+ \frac{\I g^2}{\sqrt{2}} \sum_a \Big\{ (G_{c,a} + \ol{N}{}_{c,a}) \ol{G}{}_{\Xi} - (\ol{G}{}_{c,a} + N_{c,a}) G_{\Xi} \Big\}
+ \frac{g^2}{2} \sum_{a,b} (G_{c,a} + \ol{N}{}_{c,a}) (\ol{G}{}_{c,b} + N_{c,b})
\nn \\
\ & \ \ \ \ 
+ \sum_a 
\frac{1}{e_a^2} 
\big| D_{c,a} - \sqrt{2} \, e_a^2 \, \ol{q}{}_a \ol{\wt{q}}{}_a \big|^2
- \sum_a 2 e_a^2 |q_a \wt{q}_a|^2
\nn \\
\ & \ \ \ \ 
- \sum_a D_{c,a} \Big\{ (r^1 - s^1_a) + \I (r^2 - s^2_a) \Big\}
- \sum_a \ol{D}{}_{c,a} \Big\{ (r^1 - s^1_a) - \I (r^2 - s^2_a) \Big\}
\nn \\
\ & \ \ \ \ 
- \sum_a \frac{\I}{2 e_a^2} 
\big( D_{c,a} - \sqrt{2} \, e_a^2 \, \ol{q}{}_a \ol{\wt{q}}{}_a \big)
\Big\{ (\del_0 - \del_1) \ol{B}{}_{c\+,a}
+ (\del_0 + \del_1) \ol{A}{}_{c=,a}
- \I (\del_0^2 - \del_1^2) \ol{\phi}{}_{c,a} \Big\}
\nn \\
\ & \ \ \ \ 
+ \sum_a \frac{\I}{2 e_a^2} 
\big( \ol{D}{}_{c,a} - \sqrt{2} \, e_a^2 \, q_a \wt{q}_a \big)
\Big\{ (\del_0 - \del_1) B_{c\+,a}
+ (\del_0 + \del_1) A_{c=,a}
+ \I (\del_0^2 - \del_1^2) \phi_{c,a} \Big\}
\nn \\
\ & \ \ \ \ 
+ \half \sum_a (\phi_{c,a} + \ol{\phi}{}_{c,a}) (\del_0^2 - \del_1^2) r^1 
+ \sum_a \frac{1}{4 e_a^2} \Big|
  (\del_0 - \del_1) B_{c\+,a}
+ (\del_0 + \del_1) A_{c=,a}
+ \I (\del_0^2 - \del_1^2) \phi_{c,a}
\Big|^2
\nn \\ 
\ & \ \ \ \  
- \frac{g^2}{2} \sum_a \Big\{
(B_{c\+,a} + \ol{B}{}_{c\+,a}) (\del_0 - \del_1) y^2
- (A_{c=,a} + \ol{A}{}_{c=,a}) (\del_0 + \del_1) y^2
\Big\}
\nn \\
\ & \ \ \ \ 
+ \frac{\I}{2} \sum_a 
(\phi_{c,a} - \ol{\phi}{}_{c,a}) (\del_0^2 - \del_1^2) r^2
- \frac{g^2}{2} \sum_{a,b} 
(A_{c=,a} + \ol{A}{}_{c=,a}) (B_{c\+,b} + \ol{B}{}_{c\+,b})
\nn \\
\ & \ \ \ \ 
+ \frac{\I}{2} \sum_a \Big\{
(B_{c\+,a} - \ol{B}{}_{c\+,a}) (\del_0 - \del_1) r^1
+ (A_{c=,a} - \ol{A}{}_{c=,a}) (\del_0 + \del_1) r^1 
\Big\}
\nn \\
\ & \ \ \ \ 
+ \text{(fermionic terms)}
\, . \label{GLSM-522-b1}
\end{align}
Notice that we have not imposed any gauge-fixing conditions on the Lagrangian. 
The form (\ref{GLSM-522-b1}) looks very complicated.
However, this is reduced to the following simple form under the equations of motion (\ref{EOM-aux}) and for the auxiliary fields $(D_{V,a}, D_{\Phi,a}, F_a, \wt{F}_a, G_{\Xi}, G_{\Gamma})$ in the irreducible superfields $(\Sigma_a, \Phi_a, Q_a, \wt{Q}_a, \Xi, \Gamma)$:
\begin{align}
\Scr{L}
\ &= \ 
\sum_a \frac{1}{e_a^2} \Big\{
\half (F_{01,a})^2 
- |\del_m \sigma_a|^2
- |\del_m M_{c,a}|^2
\Big\}
- \sum_a \Big\{ 
|D_m q_a|^2
+ |D_m \wt{q}_a|^2
\Big\}
\nn \\
\ & \ \ \ \ 
- \frac{1}{2 g^2} \Big\{ (\del_m r^1)^2 + (\del_m r^3)^2 \Big\}
- \frac{g^2}{2} \Big\{ (\del_m y^2)^2 + (D_m \gamma^4)^2 \Big\}
+ \sqrt{2} \, \ve^{mn} \sum_a \del_m \big( (r^4 - t^4_a) A_{n,a} \big)
\nn \\
\ & \ \ \ \ 
- 2 g^2 \sum_{a,b} \big( \sigma_a \ol{\sigma}{}_b + M_{c,a} \ol{M}{}_{c,b} \big)
- 2 \sum_a \big( |\sigma_a|^2 + |M_{c,a}|^2 \big) 
\big( |q_a|^2 + |\wt{q}_a|^2 \big)
\nn \\
\ & \ \ \ \ 
- \sum_a \frac{e_a^2}{2} \Big\{ |q_a|^2 - |\wt{q}_a|^2 - \sqrt{2} \, (r^3 - t^3_a) \Big\}^2
- \sum_a e_a^2 \Big| \sqrt{2} \, q_a \wt{q}_a + \big( (r^1 - s^1_a) + \I (r^2 - s^2_a) \big) \Big|^2
\nn \\
\ & \ \ \ \ 
+ \frac{g^2}{2} \sum_{a,b} 
(A_{c=,a} + \ol{A}{}_{c=,a}) (B_{c\+,b} + \ol{B}{}_{c\+,b})
\nn \\
\ & \ \ \ \ 
+ \text{(fermionic terms)}
\, . \label{GLSM-522-b2}
\end{align}
We have also imposed the equations of motion for non-dynamical fields $(F_{c,a}, G_{c,a}, N_{c,a})$ in the prepotential $C_a$.
We immediately find that the vectorial fields $(A_{c=,a}, B_{c\+,a})$ still contribute to the Lagrangian (\ref{GLSM-522-b2}).
Indeed they are coupled to the (non)-dynamical fields via the duality relations (\ref{r2+y2}) and (\ref{r2-y2}).
If these vectorial fields are gauged away by the complex linear superfield $\wt{C}_a = X_a + Y_a + \ol{Z}{}_a = L_a$, we cannot obtain the nongeometric structure on the target space geometry of the IR sigma model.
This is nothing but the crucial point in the previous work \cite{Kimura:2013fda}.
Thus, in order to remove genuinely redundant degrees of freedom only,
we have to introduce a supergauge parameter $\wt{C}_a$ which is not a complex linear superfield.
If we find a suitable gauge-fixing, this should keep the vectorial fields $(A_{c=,a}, B_{c\+,a})$ non-trivial.
Then we can analyze the supersymmetric vacua of (\ref{GLSM-522-b2}),
and eventually obtain the correct sigma model in the IR limit.

\end{appendix}
%%%%%%%%%%%%%%%%%%%%%%%%%%%%%%%%%%%%%%%%%%%%%%%%%%%%%%%%%%%%%%%%%%%%%%
%\newpage

}
%%%%%%%%%%%%%%%%%%%%%%%%%%%%%%%%%%%%%%%%%%%%%%%%%%%%%%%%%%%%%%%%%%%%%%
\end{document}